\documentclass[apj]{emulateapj}

\usepackage{multirow}
\usepackage{graphicx}
\usepackage{amsmath}
\usepackage{epsfig}
\usepackage{amssymb}

\usepackage{hyperref}	
\hypersetup{colorlinks=true,linkcolor=blue,citecolor=blue,filecolor=blue,urlcolor=blue}

\newcommand{\blue}[1]{\textcolor{blue}{#1}}

\begin{document}


\title{Advanced LIGO Constraints on Neutron Star Mergers and R-Process Sites}

\author{Benoit C\^ot\'e,\altaffilmark{1,2,8,9}, Krzysztof Belczynski,\altaffilmark{3} Chris L. Fryer,\altaffilmark{4,9} Christian Ritter,\altaffilmark{1,8,9} \\ Adam Paul,\altaffilmark{1} Benjamin Wehmeyer,\altaffilmark{5} Brian W. O'Shea\altaffilmark{2,6,7,8}}

\altaffiltext{1}{Department of Physics and Astronomy, University of Victoria, Victoria, BC, V8W 2Y2, Canada}
\altaffiltext{2}{National Superconducting Cyclotron Laboratory, Michigan State University, East Lansing, MI, 48824, USA}
\altaffiltext{3}{Astronomical Observatory, Warsaw University, Al. Ujazdowskie 4, 00-478 Warsaw, Poland}
\altaffiltext{4}{Computational Physics and Methods (CCS-2), LANL, Los Alamos, NM, 87545, USA}
\altaffiltext{5}{Department of Physics, University of Basel, Klingelbergstrasse 82, CH-4056 Basel, Switzerland}
\altaffiltext{6}{Department of Physics and Astronomy, Michigan State University, East Lansing, MI, 48824, USA}
\altaffiltext{7}{Department of Computational Mathematics, Science and Engineering, Michigan State University, East Lansing, MI, 48824, USA}
\altaffiltext{8}{Joint Institute for Nuclear Astrophysics - Center for the Evolution of the Elements, USA}
\altaffiltext{9}{NuGrid Collaboration, \href{http://nugridstars.org}{http://nugridstars.org}}

\label{firstpage}

\begin{abstract}
The role of compact binary mergers as the main production site of r-process elements is investigated by combining stellar abundances of Eu observed in the Milky Way, galactic chemical evolution (GCE) simulations, binary population synthesis models, and Advanced LIGO gravitational wave measurements. We compiled and reviewed seven recent GCE studies to extract the frequency of neutron star - neutron star (NS-NS) mergers that is needed in order to reproduce the observed [Eu/Fe] vs [Fe/H] relationship. We used our simple chemical evolution code to explore the impact of different analytical delay-time distribution (DTD) functions for NS-NS mergers. We then combined our metallicity-dependent population synthesis models with our chemical evolution code to bring their predictions, for both NS-NS mergers and black hole - neutron star mergers, into a GCE context. Finally, we convolved our results with the cosmic star formation history to provide a direct comparison with current and upcoming Advanced LIGO measurements. When assuming that NS-NS mergers are the exclusive r-process sites, and that the ejected r-process mass per merger event is 0.01\,M$_\odot$, the number of NS-NS mergers needed in GCE studies is about 10 times larger than what is predicted by standard population synthesis models. These two distinct fields can only be consistent with each other when assuming optimistic rates, massive NS-NS merger ejecta, and low Fe yields for massive stars. For now, population synthesis models and GCE simulations are in agreement with the current upper limit (O1) established by Advanced LIGO during their first run of observations. Upcoming measurements will provide an important constraint on the actual local NS-NS merger rate, will provide valuable insights on the plausibility of the GCE requirement, and will help to define whether or not compact binary mergers can be the dominant source of r-process elements in the Universe.
\\
\end{abstract}


\section{Introduction}
\label{sect_intro}
Understanding the production sites and the evolution
of chemical elements in the Universe represents a significant challenge that 
requires a coherent effort between multiple fields, from nuclear physics to
galaxy evolution and cosmological structure formation.  
Galactic chemical evolution (GCE, e.g., \citealt{c01,gibson03,nomoto13,matteucci14}) models
and simulations are powerful tools to address this
multidisciplinary topic, as they allow the astrophysical community to bridge the area of
nuclear astrophysics (e.g., nucleosynthetic yields) and 
the observation of stellar abundances in local galaxies.
They also provide valuable insights on how chemical species
are mixed, recycled, and dispersed inside and outside galaxies,
which ultimately leads to a better understanding of how galaxies form, evolve,
and interact with their surrounding.  In the recent years, an
effort has been made to distinguish which
site, between core-collapse supernovae (CC~SNe) and neutron star - neutron star (NS-NS)
mergers, is the dominant source of r-process elements (e.g., \citealt{a04,wi06,agt07,t11,m14,mv14,c15,w15}).
As described below, the purpose of this paper is to provide an additional
constraint on this issue.

It is generally agreed that the presence of r-process elements, typically
using the [Eu/Fe] ratio observed on the surface of extremely metal-poor
stars with [Fe/H]~$\lesssim-3$ in the Milky Way, is difficult to explain
with NS-NS mergers alone, as the delay time needed between
the formation of binary neutron stars and their coalescence prevents them
from appearing before [Fe/H]~$\sim-2.5$ (e.g., \citealt{a04,m14,c15,w15}).
This suggested that CC~SNe could be an important r-process source
in the early Universe, as they would eject r-process elements and other lighter metals,
such as Fe, simultaneously.  However, in the last two years, more sophisticated studies using semi-analytical models and
cosmological hydrodynamic simulations have shown that a reduced star formation
efficiency in low-mass progenitor galaxies can 
slow down the early evolution of [Fe/H], which in turn allows NS-NS mergers to
appear at [Fe/H]~$<-3$ (e.g., \citealt{h15,ks16}).  This has also been seen in the one-zone models
of \cite{i15}.  Furthermore, \cite{s15} showed that non-uniform mixing in
zoom-in cosmological hydrodynamic simulations, which generates scatter
in the age$-$metallicity relationship, can push the appearance of 
NS-NS mergers to lower [Fe/H] values compared to simpler models (see also \citealt{vv15}).

As a summary, it appears that NS-NS mergers alone could explain the evolution of r-process 
elements when using a proper treatment of how galaxies assemble and how
metals mix with primordial gas in the early Universe.  It is therefore difficult to evaluate the relative
contribution of CC~SNe and NS-NS mergers based only on the appearance of NS-NS mergers on the
[Fe/H] axis.  In addition, in order to validate the conclusions
drawn by GCE studies, it is important to combine numerical predictions with
additional constraints that are not directly related to chemical evolution.  Otherwise,
when stellar abundances are the only constraint, it can be difficult to
distinguish between reliable and misleading conclusions (see \citealt{c16a}).
Here are a few examples of additional observations that can be used by GCE studies
-- gas content (e.g, \citealt{kubryk15}), galactic inflow and outflow rates (e.g., \citealt{martin99,lehner11}),
star formation efficiencies (e.g., \citealt{leroy08}), and galactic structure and dynamics (e.g., \citealt{minchev13,minchev14}).

To constrain NS-NS mergers, the merger rates adopted in GCE studies can be compared
to the merger rates predicted by population synthesis models (e.g., \citealt{d12})
and by observations of short gamma-ray bursts (e.g., \citealt{abadie10}).  In addition,
the LIGO (e.g., \citealt{a09}) and Advanced LIGO (e.g., \citealt{aasi15}) gravitational wave detectors have recently opened a new
window for improving our understanding of compact binary mergers.  The coalescence of two black
holes has been detected (\citealt{aaa16}) and measurements are currently ongoing to derive the rate
of NS-NS mergers and black hole - neutron star (BH-NS) mergers (\citealt{aaa16b}).  
These upcoming measurement will provide solid constraints that, combined with chemical evolution
models and simulations, will help defining whether or not NS-NS mergers can be the dominant
source of r-process elements in the Milky Way and in its satellite galaxies.  The LIGO horizon
extends up to a redshift of 0.7 for the most massive black hole - black hole (BH-BH) mergers.
However, for NS-NS and BH-NS mergers, we can expect a reduced horizon.
We also note that merger rates can be derived from geological $^{244}$Pu abundances (\citealt{hoto15}). 
The mass ejected by NS-NS mergers can be constrained by relativistic hydrodynamic
simulations (see Section~\ref{sect_ej_NSMs}) and from the scatter observed in the
abundances of extremely metal-poor stars (\citealt{mr16}).

In this paper, we aim to improve our understanding regarding the role of compact binary mergers 
in the production of the r-process by creating connections between GCE, population synthesis
models, and Advanced LIGO measurements.  Using interdisciplinary constraints and accounting
for various sources of uncertainties, we systematically compare seven recent chemical evolution
studies, which range from simple one-zone models to cosmological hydrodynamic zoom-in simulations,
and homogenize the number of NS-NS mergers required to reproduce the observed abundances of Eu in the Milky Way.
We also use our own simple chemical evolution model to explore different delay-time
distribution (DTD) functions, including the metallicity-dependent ones predicted by population synthesis
models, to see how they impact our ability to reproduce observations.  We provide GCE predictions
for both NS-NS and BH-NS mergers.  The ultimate goal is to determine whether or not the binary merger
rates needed in GCE simulations are reasonable and consistent with the upcoming Advanced 
LIGO measurements and population synthesis predictions.

This paper is organized as follows.  In Section~\ref{sect_ibm}, we present the basic ingredients for
implementing compact binary mergers in chemical evolution studies.  In Section~\ref{sect_GCE}, we
present seven recent GCE studies found in the literature, describe our own chemical evolution code,
and provide an normalized comparison for the needed number of NS-NS mergers per unit of
stellar mass formed.  We explore, in Section~\ref{sect_DTD}, the impact of different analytical DTD functions
for NS-NS mergers.  In Section~\ref{sect_psm}, we present our population synthesis models and include their
predicted DTD functions and merger frequencies, for NS-NS and BH-NS mergers, into our chemical evolution code
as an input.  We convolved, in Section~\ref{sect_cmr}, the merger rates and frequencies needed in GCE studies and
predicted by population synthesis models with the cosmic star formation history to allow a direct
comparison with Advanced LIGO measurements.  Our discussion and conclusions are presented in
Sections~\ref{sect_disc} and \ref{sect_concl}, respectively.

\section{Implementation of Binary Mergers}
\label{sect_ibm}
In the following sections, we describe how to implement r-process production sites in GCE studies (see also \citealt{m14,c15}).  The approach is similar to the implementation of Type~Ia SNe and can be used for NS-NS and BH-NS mergers and for CC~SNe.  However, because of the purpose of this paper, we mainly focus on compact-binary mergers.

\subsection{Total Mass Ejected}
\label{sect_me}
The total mass ejected by each binary merger allows us
to scale the abundance pattern associated with the r-process and
to determine the mass of heavy elements returned into the interstellar medium.  

\subsubsection{Neutron Star Mergers}
\label{sect_ej_NSMs}
Table~\ref{tab_m_ej} shows a compilation of the wide range of total mass ejected by NS-NS mergers,
which is typically predicted by relativistic hydrodynamic simulations.
In the models of \cite{korobkin12}, the ejected masses
range from $\sim$\,0.008 to $\sim$\,0.04\,M$_\odot$, but the most massive ejecta only
occurs in peculiar progenitor mass ratios.  In addition, the Newtonian smooth particle hydrodynamics simulations of
\cite{korobkin12} possibly predict more ejecta than other calculations due to the lack of relativistic
effects (\citealt{bauswein13}). \cite{fryer15} used the ejected masses found in \cite{korobkin12} and included them into
population synthesis calculations.  They found that although the upper limit for the ejected mass is around 0.035\,M$_\odot$, the vast
majority of NS-NS mergers should eject less than 0.02\,M$_\odot$.

\begin{deluxetable}{cc}
\tablewidth{0pc}
\tablecaption{Compilation of Predicted Total Ejected Masses for Neutron Star Mergers\label{tab_m_ej}}
\tablehead{ \colhead{Reference} & \colhead{Ejected Mass [$10^{-2}$ M$_\odot$]} }
\startdata
\cite{korobkin12} & $0.76 - 3.9^{a}$ \\
\cite{bauswein13} & $0.1 - 1.5$ \\
\cite{hotokezaka13} & $0.03 - 1.4$ \\
\cite{fryer15} & $1.0 - 3.5^{a}$ \\
\cite{endrizzi16} & $<0.1 - 1.0$ \\
\cite{radice16} & $0.02 - 12.5^{a}$ \\
\cite{sekiguchi16} & $0.2 - 1.3$ 
\enddata
\tablenotetext{a}{See text for discussion.}
\end{deluxetable}

In addition to the dynamical ejecta mentioned above, outflows from the accretion disc of NS-NS mergers
can also eject a considerable amount of material (e.g., \citealt{fm13,fkmq15,just15}).  Although these outflows 
can be composed of r-process elements (e.g., \citealt{just15,wu16}), it is generally believed that disc outflows
have a higher electron fraction (less neutron rich) than the dynamical ejecta.  Several nucleosynthesis models
argue that such ejecta only produces elements with atomic weights below 120-130 (e.g., \citealt{perego14,martin15})
and would therefore not contribute significantly to the evolution of Eu in our chemical evolution model.

One proposal to increase the ejecta of NS-NS mergers is to consider collisions instead of mergers,
as the mass ejected in the case of highly eccentric binaries can be an order of magnitude larger than in the case of circular orbits.  
\cite{radice16}, in their simulations, assumed parabolic orbits and ranged the periastron of these orbits to determine the mass of the ejecta.  These parabolic orbits
are produced in dense clusters with close encounters of compact objects, which are typically described in terms
of impact separations (impact parameter) and velocities.  In their simulations, the most massive ejecta 
only occurred for eccentric orbits with periastron distances below $\sim$~15\,km.  However, dynamical
interactions producing such close orbits are rare and, overall, these systems are not
expected to contribute significantly to the production of r-process material (see Appendix~\ref{ap_1}).

Attempts to observe macronova candidates have discovered two possible candidates: GRB 130603B (\citealt{tanvir13,berger13}) and GRB 060614 (\citealt{jin15,yang15}). However, the nearby short burst GRB 150101B (\citealt{fong16}) places upper limits on the near infrared emission below these purported detections. On top of this, recent calculations of the opacities of the r-process elements produced in r-process ejecta argue that this ejecta is not bright in the near-IR (\citealt{fontes15}). These opacity arguments suggest that the emission observed from GRB 130603B and GRB 060614 cannot be from heavy r-process ejecta. Either the emission of these two bursts arises from wind ejecta (and hence does not measure the r-process ejecta) or these are false detections. In either case, there is no longer any evidence for dynamical ejecta masses above those produced in simulations.

The total mass ejected by NS-NS mergers seems to converge toward a value that is below 0.02\,M$_\odot$, but
the range is still very uncertain.  To highlight this uncertainty, we consider in this study that NS-NS mergers
can eject, on average, a total mass between 0.001 and 0.025\,M$_\odot$, with a fiducial value of 0.01\,M$_\odot$.

\subsubsection{Black Hole - Neutron Star Mergers}
\label{sect_ej_BH_NSMs}
BH-NS mergers could potentially produce more ejecta than NS-NS mergers.  The
work of \cite{kawaguchi15} suggests that the ejected mass can range between
less than 10$^{-4}$ up to $7.9\times10^{-2}$\,M$_\odot$.  The total mass ejected
by BH-NS mergers is sensitive to the spin of the companion black hole.  According
to \cite{bauswein14}, the ejecta from a binary merger involving a non-rotating black hole
ranges from $2\times10^{-6}$ to $4\times10^{-4}$\,M$_\odot$,
while the ejecta is increased up to $8.6\times10^{-2}$ and $9.6\times10^{-2}$\,M$_\odot$
with fast rotating black holes.  In addition, the mass ejected by BH-NS mergers also depends on the 
still unknown equation of state of neutron stars (\citealt{hotokezaka13b}).  However, the rates of BH-NS mergers are typically lower
than NS-NS mergers (see Section~\ref{sect_psm}).  

Having or not a mass ejection in BH-NS mergers is determined by orbital parameters at merger.
In particular, if the tidal disruption distance between a neutron star and a black hole is smaller than
the black hole event horizon, no mass ejection is expected. The extent of the black hole horizon depends
sensitively on the spin of the black hole.  The neutron star tidal disruption radius depends on the equation of state,
the BH-NS mass ratio, and on the inclination of the neutron star's orbit with respect to the black hole
spin. It was estimated that for highly spinning black holes, about $40\%$ of BH-NS mergers
lead to the disruption of the neutron star outside the black hole horizon, and therefore to mass
ejection.  However, for slow spinning black holes, neutron star disruptions outside the horizon is
only expected to occur in $\sim\,2\%$ of the BH-NS mergers (\citealt{Belczynski2008b}).

\subsection{R-Process Yields}
Once the total mass ejected per r-process event is known, yields for heavy elements need to
be selected in order to calculate the ejected mass of individual elements.
Theoretical r-process nucleosynthetic calculations 
are making progress for compact binary mergers (\citealt{korobkin12,rosswog14,lippuner15,just15,roberts16}) and CC~SNe
(e.g., \citealt{am11,arcones13,nakamura15,nishimura15}).
Those yields can be used in chemical evolution models to probe 
the relative contribution of different production sites.
However, those nucleosynthetic calculations, especially for Eu, are subject to many
sources of uncertainties which are currently affecting the reliability of GCE
predictions.  Among others, nuclear masses can induce significant uncertainties in
the predicted yields (\citealt{m16,mum16}) while different fission models can change
their abundance patterns (\citealt{korobkin12,g15,e16}).

To reduce the uncertainties in this work, we adopt the observed r-process yields 
extracted using the solar r-process residuals method (\citealt{agt07}), which 
implies a Eu mass fraction of $10^{-3}$ for each r-process event.  Although
the use of empirical yields prevents us from drawing conclusions on how the
r-process elements are made, we believe it is a good choice for the purpose of this paper.
Our goal is to investigate whether or not NS-NS and BH-NS mergers are frequent 
enough to be able to reproduce the abundances of Eu observed in stars, assuming
that compact binary mergers generate the r-process abundances pattern.

\subsection{Delay-Time Distribution Function}
The DTD function defines how many r-process events occur as a function of time in a given stellar population.
If the r-process originates from CC~SNe, the DTD function is therefore linked to the SN
rate, which is defined by the initial mass function (IMF) and the lifetime of massive stars.  If the 
r-process originates from compact binary mergers, the DTD function accounts for the delay 
needed between star formation, the formation of neutron stars and black
holes, and the coalescence timescale of compact binary systems.  In this last case, the DTD function can
be seen as the probability for a binary system to merger as a function of time.  In all cases,
the merger rates are affected by the assumed  binary fraction among the massive stars.
As described in the next sections, many forms of DTD function can be used in GCE studies, from simple
analytical prescriptions to DTD functions predicted by population synthesis models (e.g., \citealt{d12}).

As for the total mass ejected per r-process event, the normalisation of the DTD functions
allows to scale the total amount of material returned into the interstellar medium.  In our work,
this normalisation parameter is described in units of number of r-process events per
stellar mass formed, which represents the total number of NS-NS or BH-NS mergers that will
occur during the lifetime of a stellar population.  

\section{Galactic Chemical Evolution Studies}
\label{sect_GCE}
Several GCE studies, which are reviewed in the next sections, have recently been conducted to determine if
NS-NS mergers can be the dominant source of r-process in the Universe.  Those studies usually aim
to reproduce the abundance evolution of Eu (and sometimes other heavy elements, \citealt{a04,c15,i15,ks16}) observed in
the Milky Way.  Conclusions are typically based on whether or not NS-NS mergers can occur early
enough to explain the evolution of our Galaxy at low [Fe/H].  However, as explained in 
Section~\ref{sect_intro}, the evolution of Eu in the early Universe is sensitive to 
the choice of code implementation and on how mass assembly is treated.  In the present 
study, we complement the previous works by investigating the role of NS-NS mergers based on
the total number of merger events needed in GCE simulations.  As shown in
Sections~\ref{sect_psm} and \ref{sect_cmr}, this number can directly be compared
with population synthesis models and Advanced LIGO measurements.

In the following sections, we briefly describe the chemical evolution codes used 
in different studies found in the literature.  The extracted total number of NS-NS mergers 
per stellar mass formed, for all studies including ours, are shown in the upper
panel of Figure~\ref{fig_EuFe}.  To provide a more reliable comparison
between the different studies (see Section~\ref{sect_nc}), we also show, in the middle
and lower panels, the mass of Eu ejected per NS-NS merger event and the mass of Fe ejected
by massive stars per stellar mass formed.  The work of \citealt{vv15} included
neutron star mergers, but only considered the total ejected mass of r-process material.
Therefore, in order to provide a consistent comparison among our selection of studies, which
includes a normalization of Eu yields (see Section~\ref{sect_nc}), we decided to
exclude this study from our sample.  The work of \cite{vv15}, however, is important
for the interpretation of our results and is considered in Section~\ref{sect_pldtd}.

To be consistent with previous studies, we first focus on NS-NS mergers, but we include
the contribution of BH-NS mergers in Section~\ref{sect_psm}. We note that, for each selected
study, we only considered the input parameters that generated the best agreement
with observations.  We recall that some of the studies reviewed in the following
sections did include CC~SNe as well as NS-NS mergers as sources of r-process elements.
For those studies, we only considered the cases where NS-NS mergers were the exclusive
r-process site.  In several studies, which include ours, the binary fraction for massive
stars is not specified, as it is indirectly included in the normalization parameters
regulating the merger rates.

\subsection{Matteucci et al. (2014)}
\citet[M14]{m14} used a recent version of the two-infall model originally developed
by \cite{c97,c01}.  It consists of a multi-zone model where the star formation rate
is self-generated across the Galactic radius and calculated from a gas density profile,
which evolves over time.  The choice of stellar yields is the same as in the model 15 of \cite{r10}.

\blue{M14} assume that each NS-NS merger event ejects $3\times10^{-6}$\,M$_\odot$ of Eu. For
each stellar population, the rate of NS-NS mergers follows the rate of appearance of neutron stars,
which is linked to the lifetimes of massive stars.  A constant delay is then added to the
NS-NS merger rate in order to account for the time needed for neutron star binaries to coalesce.  They adopted the IMF of \cite{s86} with a lower\footnote{We note that we found the lower limit in \cite{r10}.}
and upper mass boundaries of 0.1 and 100\,M$_\odot$.  A fraction of 1.8\,\% of all stars with initial mass between 9 and 30\,M$_\odot$
are assumed to be the progenitors of NS-NS mergers.  With this setup, we derived a total number of 
$4.34\times10^{-5}$ NS-NS merger per unit of stellar mass formed.

The stellar yields of massive stars are taken from \citet[K06]{k06}, which includes 7 masses between 13
and 40\,M$_\odot$ and 4 metallicities.  All stars with initial mass equal or above 20\,M$_\odot$
undergo a hypernova.  According to \cite{r10}, massive star yields are applied to stars in the mass
range of 8 to 100\,M$_\odot$.  All stars with an initial mass above 40\,M$_\odot$ are assumed
to eject the same material than the 40\,M$_\odot$ models.  A linear interpolation is performed
between 6 and 13\,M$_\odot$, where 6\,M$_\odot$ represents the most massive asymptotic
giant branch (AGB) models of \cite{karakas10}.  After interpolating the models of \cite{karakas10}
to provide the same metallicities as in \blue{K06}, we found that $2.84\times10^{-4}$\,M$_\odot$ of
Fe is ejected in total per unit of stellar mass formed.  We associated the lowest metallicity of
provided by \cite{karakas10}, $Z=0.0001$ in mass fraction, to the zero-metallicity yields of \blue{K06},
which yielded similar results as when assuming no Fe ejecta for AGB stars at $Z=0$.

\subsection{Cescutti et al. (2015)}
\citet[C15]{c15} used an inhomogeneous chemical evolution model that originates
from the work of \cite{c08} and \cite{chiap08}, who aimed to reproduce metal-poor
stars.  In order to follow the stochasticity at early time, the Galactic halo has been 
split into several independent cells in which the IMF was randomly sampled, following
an infall-driven star formation rate.

On average, each NS-NS merger is assumed to eject $5\times10^{-6}$\,M$_\odot$ of Eu.  \blue{C15} considered
that 2\,\% of all massive stars are part of binary systems that eventually produce a NS-NS merger.  As in
\blue{M14}, the rate of NS-NS mergers follows the rate of CC~SNe, but is delayed by a constant coalescence
timescale.  The adopted IMF is the one of \cite{s86} with a lower and upper mass boundaries of 0.1 and 100\,M$_\odot$.
With a minimum mass of 8\,M$_\odot$ for massive stars, we have calculated that $3.56\times10^{-5}$
NS-NS merger occur per unit of stellar mass formed.  The core-collapse Fe yields presented in \cite{f04} are applied to 
all stars between 8 and 100\,M$_\odot$.  From the IMF properties, we found that massive stars eject 
$3.83\times10^{-4}$\,M$_\odot$ of Fe per stellar mass formed.

\subsection{Hirai et al. (2015)}
\citet[H15]{h15} performed a series of hydrodynamic simulations of dwarf spheroidal galaxies,
using the ASURA code described in \cite{s08,s09}, to represent the building blocks
of the Galactic halo.

The mass of Eu ejected by each NS-NS merger has been set to $2\times10^{-5}$\,M$_\odot$ (private communication) in
order to reach the desired level of [Eu/Fe]\,$=0.5$ at [Fe/H]\,$=0.0$.  We note that
\blue{H15} did not include Type~Ia SNe (SNe~Ia) in their simulations, as the goal of their paper was to address
the Eu abundances observed in extremely metal-poor stars (at [Fe/H]\,$\lesssim-3$).
The IMF of \cite{s55} is considered with a lower and upper mass boundaries of 0.1
and 100\,M$_\odot$.  With the assumption that 1\,\% of all stars in the mass range
between 8 and 20\,M$_\odot$ produce NS-NS mergers, we derive a total of $5.45\times10^{-5}$
NS-NS merger per unit of stellar mass formed.  The stellar yields of \blue{K06} are applied to
stars with an initial mass between 8 and 40\,M$_\odot$.  With the IMF used in \blue{H15},
we calculated that, on average, $5.71\times10^{-4}$\,M$_\odot$ of Fe is ejected
per unit of stellar mass formed.

\begin{figure}
\begin{center}
\includegraphics[width=3.25in]{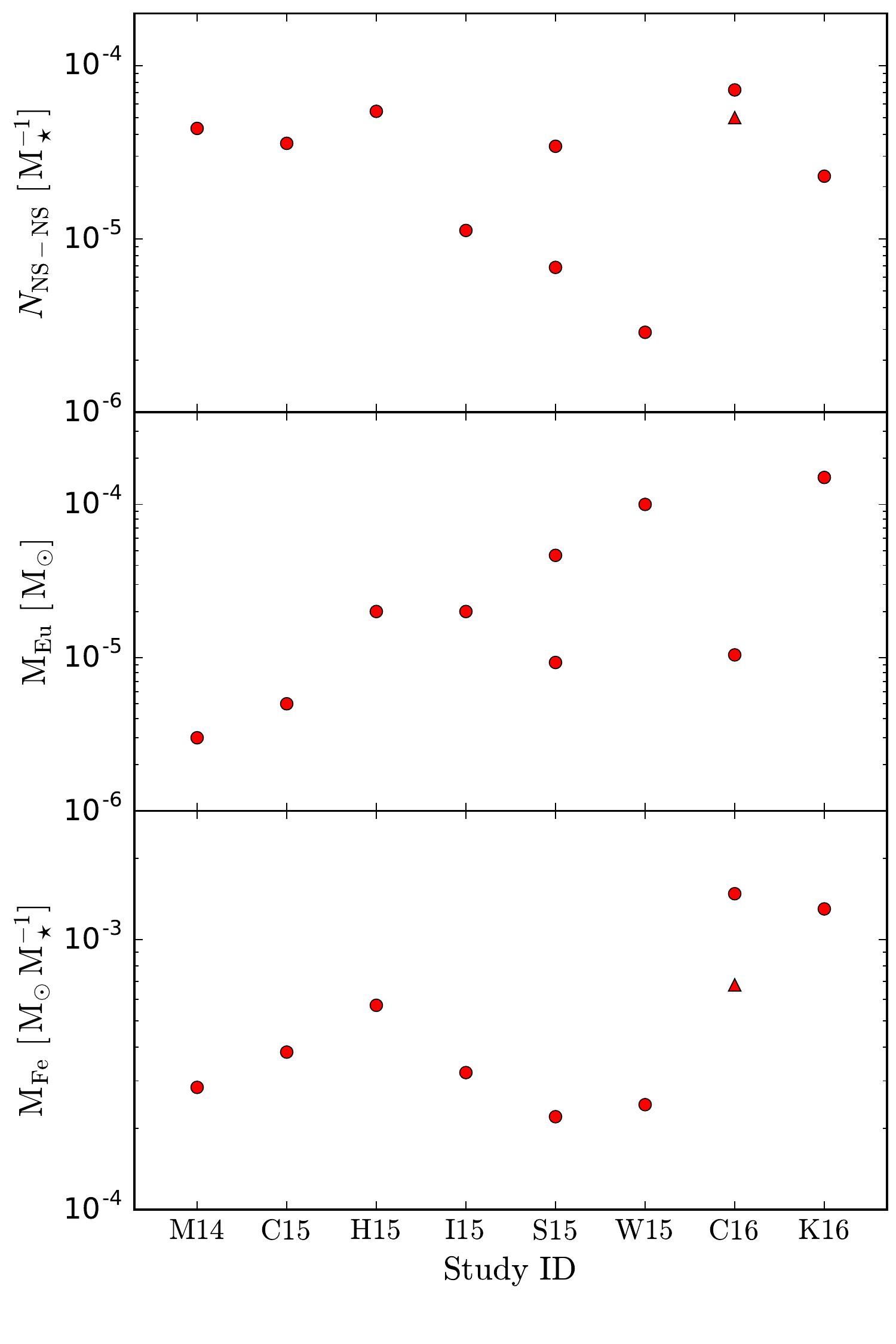}
\caption{Number of NS-NS mergers (upper panel), per unit of stellar mass formed, extracted from different chemical evolution studies.  The middle and lower panels show, respectively,
the adopted mass of Eu ejected per NS-NS merger event and the average mass of Fe ejected by massive stars per stellar mass formed.  Our study is labeled as C16.  The other studies
taken from the literature are \protect\citet[M14]{m14}, \protect\citet[C15]{c15}, \protect\citet[H15]{h15}, \protect\citet[I15]{i15}, \protect\citet[S15]{s15}, \protect\citet[W15]{w15},
and \protect\citet[K16]{ks16}.  In our study (C16), the triangles and circles show, respectively, our results using the stellar yields calculated by \protect\cite{k06} and by West \& Heger (in preparation) and NuGrid.}
\label{fig_EuFe}
\end{center}
\end{figure}

\subsection{Ishimaru et al. (2015)}
\citet[I15]{i15} used a one-zone model based on the originally work of \cite{iw99} and \cite{i04},
which assumes homogeneous mixing.  It has been designed to follow the chemical evolution
of the building block galaxies that should have formed the Galactic halo.  Several star formation
and galactic outflow efficiencies have been explored in \blue{I15}.

In their simulations, each NS-NS merger ejects $2\times10^{-5}$\,M$_\odot$ of Eu.  In total, there are 
500 times more core-collapse SNe than NS-NS mergers.  According to \cite{iw99},
the adopted IMF is the one of \cite{s55} with a lower and upper mass boundaries of 0.05 and 60\,M$_\odot$.
Stars with initial mass between 8 and 10\,M$_\odot$ do not eject Fe, but are still expected to generate
a core-collapse SN.  From this setup, we find that $1.12\times10^{-5}$ NS-NS merger occur per unit of 
stellar mass formed.  This number, however, does not exactly match the cumulated number of
NS-NS mergers presented in the Figure~1a of \blue{I15}.  Our calculation only becomes consistent if we assume
that the final stellar masses given in \blue{I15} are corrected for stellar ejecta.  If our assumption is
valid, our value is in agreement when we assume that $\sim40$\,\% of the mass of a stellar population
is returned into the interstellar medium.  An alternative could be that \blue{I15} do not use the same IMF 
properties than what is described in \cite{iw99}.  We therefore cannot confirm that our derivation
for the number of NS-NS mergers is correct for this study.

\blue{I15} used the stellar yields of \blue{K06} for stars more massive than 10\,M$_\odot$.  Following the
IMF described above, we found that $3.22\times10^{-4}$\,M$_\odot$ of Fe is ejected on average
per unit of stellar mass formed.  We note that extrapolating the models of \blue{K06} to cover the mass
regime between 10 and 13\,M$_\odot$ generate a similar result than simply applying the yields
of the 13\,M$_\odot$ model to all stars in that mass regime.

\subsection{Shen et al. (2015)}
\citet[S15]{s15} post-processed the Eris cosmological hydrodynamic zoom-in simulation (\citealt{g11})
in order to include the contribution of NS-NS mergers.  They assumed that each NS-NS merger ejecta has a Eu mass
fraction of $9.3\times10^{-4}$, following \cite{scg08}.  With an r-process ejected mass of either 0.01 or 0.05\,M$_\odot$,
each NS-NS merger event ejects $9.3\times10^{-6}$ or $4.65\times10^{-5}$\,M$_\odot$ of Eu, respectively.
In total, $1.88\times10^6$ and $3.76\times10^5$ NS-NS mergers, depending to the
adopted NS-NS merger ejected mass, are required in order to reproduce the solar Eu to O abundance ratio.
In order to recover the number NS-NS mergers per unit of stellar mass formed, we first needed to
recover the total amount of stars formed throughout the Eris simulation.

At redshift zero, according to \cite{g11}, the stellar mass is $3.9\times10^{10}$\,M$_\odot$, which includes
the stellar remnants. The relation between the initial stellar mass and the total ejected mass for massive stars is taken 
from \cite{r96}, while the one for stars between 1 and 8\,M$_\odot$
is extracted from \cite{w87}.  Using the IMF of \cite{k93} with a lower and upper mass boundaries of 0.08 and 100\,M$_\odot$ (\citealt{r96}),
and the assumption that only stars between 1 and 40\,M$_\odot$ contribute to the stellar ejecta (\citealt{s06}),
we calculate that 29.1\,\% of the mass of a stellar population is returned into the interstellar medium.
If we assume that the bulk of the stellar mass at redshift zero is composed of
evolved stellar populations, $5.49\times10^{10}$\,M$_\odot$ of stars should be formed in total during the
simulation.  The number of NS-NS mergers per unit of stellar mass formed should then be $3.42\times10^{-5}$ when
NS-NS mergers eject 0.01\,M$_\odot$ of r-process material, and $6.85\times10^{-6}$ when they eject
0.05\,M$_\odot$. 

For the Fe yields of massive stars, \blue{S15} use the relation found in \citealt{r96} which is derived
from the models of \cite{ww95}.  Using the same IMF properties described above, we found 
that $2.21\times10^{-4}$\,M$_\odot$ of Fe is ejected per unit of stellar mass formed.

\subsection{Wehmeyer et al. (2015)}
\citet[W15]{w15} used an inhomogeneous chemical evolution model originally 
designed by \cite{a04}.  A (2\,kpc)$^3$ simulation cube representing a portion 
of the Galaxy is split into 64 000 cells where stars can
form to produce a blast-wave that pollutes neighbouring cells with heavy elements.
When a star-forming region is selected, the mass of the stars are chosen randomly
according to the adopted IMF.

Each NS-NS merger event is assumed to eject $10^{-4}$\,M$_\odot$ of Eu.  The number
of NS-NS mergers that occur during the simulation is defined by $P_\mathrm{NSM}=4\times10^{-4}$,
which represents the probability of a newly born massive star to be in a binary system leading
to a NS-NS merger event (or the frequency of NS-NS merger per number of massive star formed).  Once the progenitor stars have undergone a CC~SN and
left behind a neutron star, the binary systems merge after one fixed, constant coalescence time.
\blue{W15} uses the IMF of \cite{s55} with a lower and upper mass boundaries of 0.1 and 50\,M$_\odot$.
Given a minimum initial mass of 8\,M$_\odot$ for massive stars, there is then in total
$2.89\times10^{-6}$ NS-NS merger per unit of stellar mass formed.

The stellar yields of massive stars are taken from the models of \cite{t96} and \cite{n97}
and are applied to stars with initial mass between 10 and 50\,M$_\odot$.
In total, according to our calculation, $2.45\times10^{-4}$\,M$_\odot$ of Fe is ejected
per unit of stellar mass formed, following the IMF described above.

\begin{figure*}
\begin{center}
\includegraphics[width=7.in]{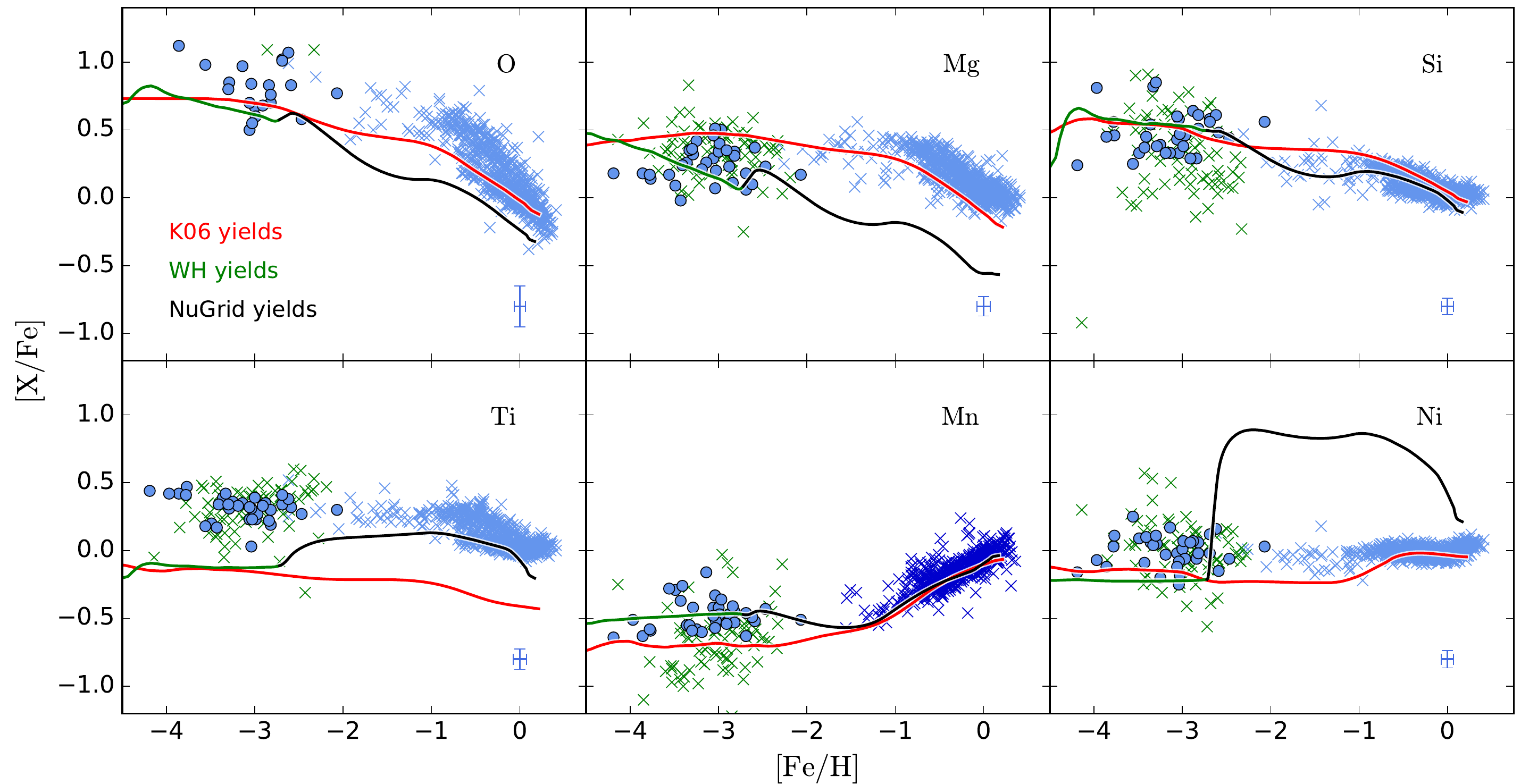}
\caption{Predicted chemical evolution of six light elements using the yields calculated by \protect\citet[red lines]{k06}, West \& Heger (in preparation, green lines), and NuGrid (black lines).
Observational data come from \protect\citet[light blue circles]{c04}, \protect\citet[green crosses]{cohen13}, \protect\citet[light blue crosses]{bfo14}, and \protect\citet[dark blue crosses]{bb15}.
The error bar symbol on the lower-right corner of each panel represents the average error of all data.}
\label{fig_light}
\end{center}
\end{figure*}

\begin{figure}
\begin{center}
\includegraphics[width=3.25in]{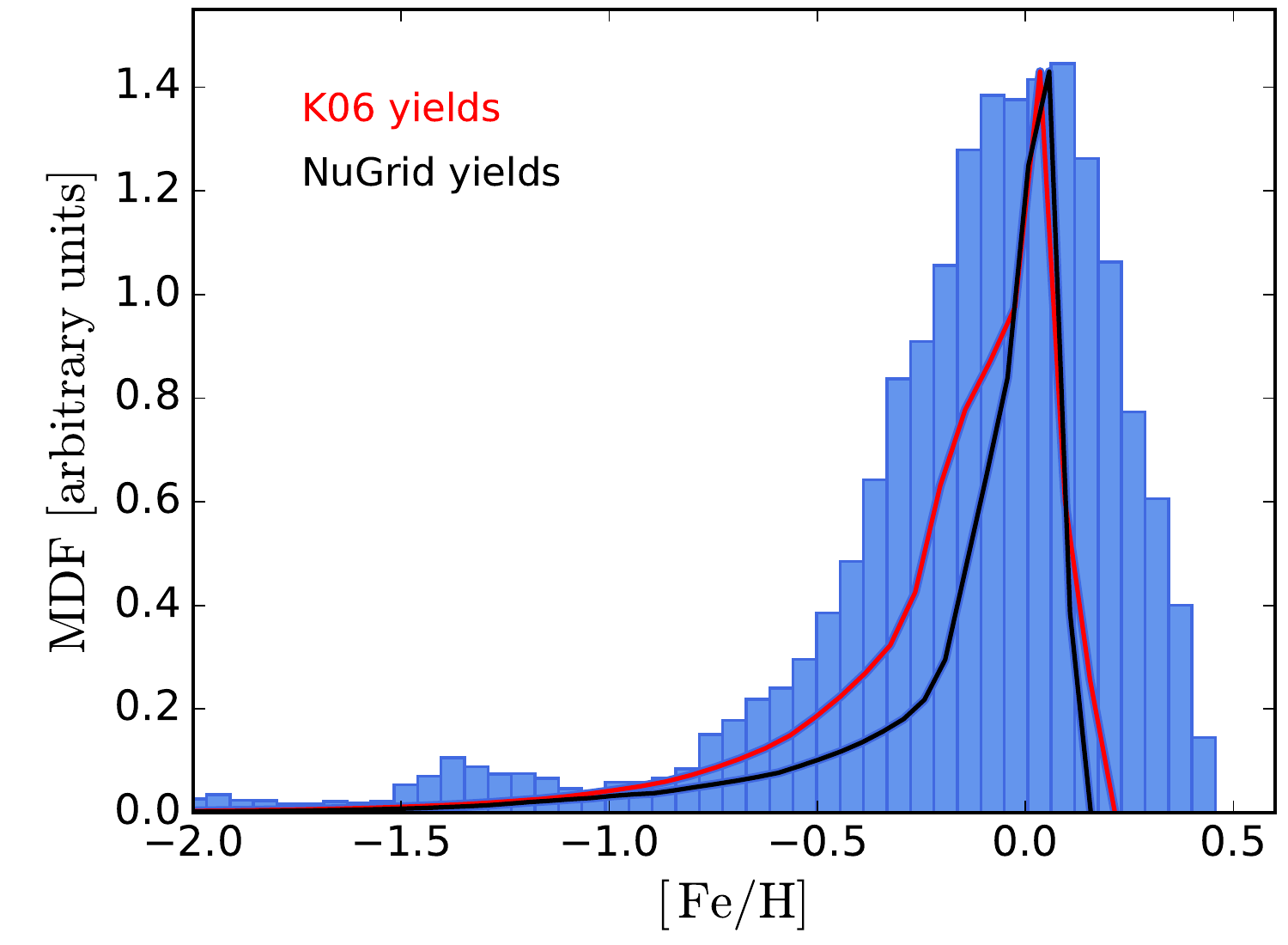}
\caption{Predicted metallicity distribution function using the yields calculated by \protect\citet[red line]{k06} and NuGrid (black line).
The blue histogram has been extracted from APOGEE R12 data and the ASPCAP reduction pipeline (\protect\citealt{gp16}).}
\label{fig_mdf}
\end{center}
\end{figure}

\begin{figure}
\begin{center}
\includegraphics[width=3.25in]{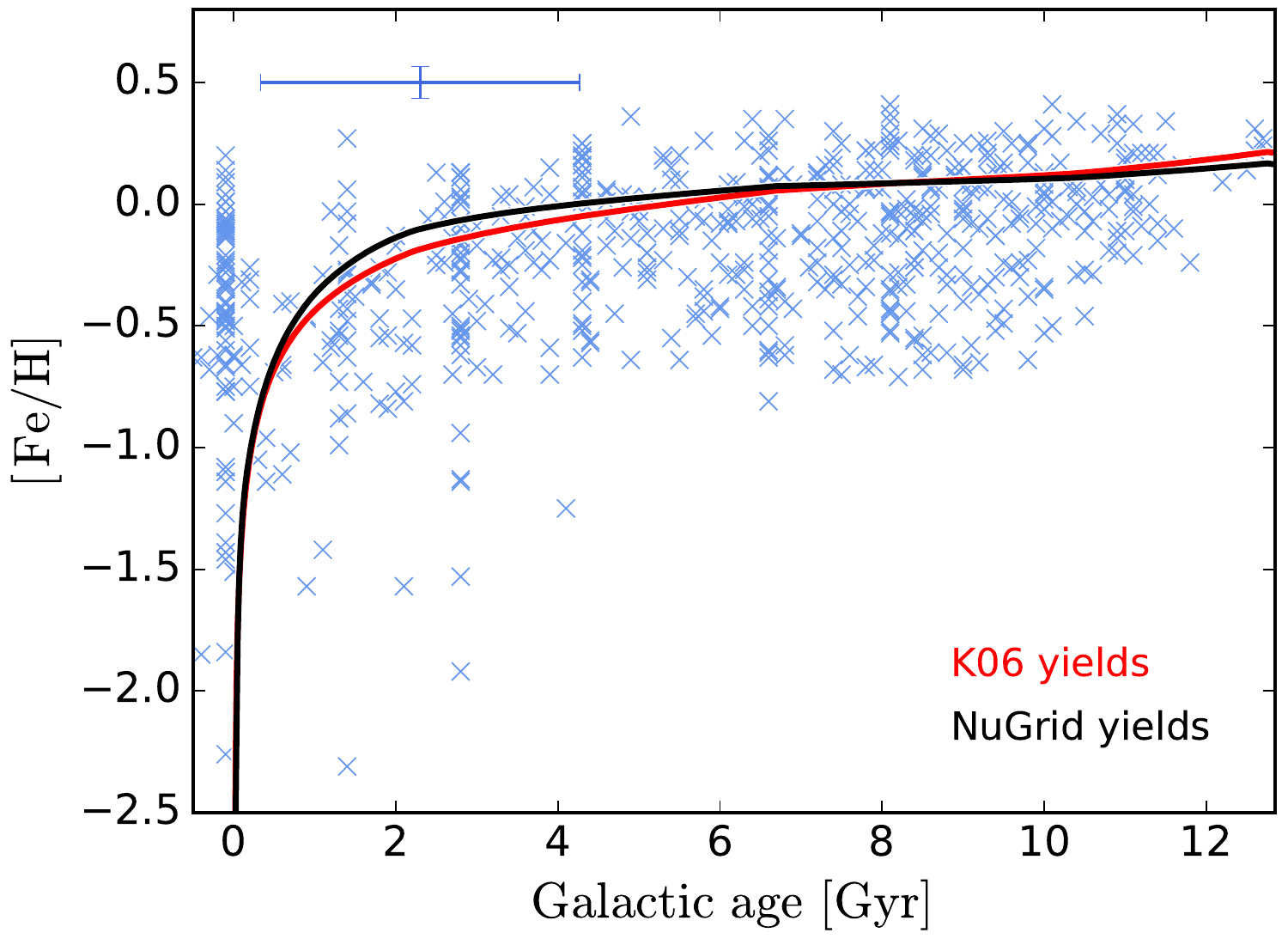}
\caption{Predicted evolution of [Fe/H] as a function of Galactic age using the yields calculated by \protect\citet[red line]{k06} and NuGrid (black line).
Observational data come from \protect\cite[blue crosses]{bfo14} where the error bar symbol on the top-left corner represents the average error of 
all data.  The error in the age measurements is typically larger for older stars.  Although
our one-zone model cannot distinguish between the different Galactic structures, the observational data include halo and disk stars.}
\label{fig_age_met}
\end{center}
\end{figure}

\subsection{Komiya \& Shigeyama (2016)}
\citet[K16]{ks16} used a semi-analytical model of galaxy formation and
evolution, described in \cite{k14}, to follow in a cosmological
context the evolution of Eu in the Milky Way.  Their model considers the
impact of galactic outflows and high-velocity NS-NS merger ejecta on the pre-enrichment
of neighbouring proto-galaxies, which eventually become part of the Milky
Way.  In each stellar population, NS-NS mergers appear following a delay-time
distribution function in the form of $t^{-1}$ with a lower and upper time
boundaries of 10\,Myr and 10\,Gyr, a choice motivated by \cite{d12}.

They assume that each NS-NS merger ejects $1.5\times10^{-4}$\,M$_\odot$ of Eu.
The adopted IMF (see their equation~2) has a similar shape than the one of \cite{c03}, but with 
different parameters to better reproduce the metallicity distribution function
of metal-poor stars.  The lower and upper mass boundaries are 0.08 and 100\,M$_\odot$,
where only stars in the mass range between 8 and 25\,M$_\odot$ are candidate progenitors
for NS-NS mergers.  With a binary fraction of 50\,\% with a flat mass ratio distribution for
primary and secondary stars, and the assumption that only 1\,\% of the
candidate binaries produce NS-NS mergers, there is $2.30\times10^{-5}$ NS-NS merger occuring
per unit of stellar mass formed (private communication).  For core-collapse SNe, the stellar yields
of \blue{K06} are applied to stars with initial mass between 10 and 40\,M$_\odot$
with a hypernova fraction of 0.5 for stars more massive than 20\,M$_\odot$ (\citealt{k14}).
With their IMF, we calculated that $1.26\times10^{-3}$\,M$_\odot$ of Fe is ejected
on average by massive stars.

\subsection{This Study}
\label{sect_omega}
{\tt OMEGA}\footnote{\href{https://github.com/NuGrid/NuPyCEE}{https://github.com/NuGrid/NuPyCEE}}
is a one-zone chemical evolution model that assumes uniform mixing and includes several prescriptions
for galactic inflows and outflows (\citealt{c16a}).  It uses an input star formation history,
along with different star formation efficiency parameterizations, to mimic the evolution
of observed and simulated galaxies.  Each stellar population represents a complete sample of the
IMF.

For massive stars, we test the yields found in \blue{K06} as well as the ones calculated by
NuGrid\footnote{\href{http://nugridstars.org/data-and-software/yields/set-1}{http://nugridstars.org/data-and-software/yields/set-1}}
 (\citealt{p13}; C. Ritter et al. in preparation) with the delayed remnant mass prescription
described in \citealt{fbw12}.  We supplement NuGrid yields with the zero-metallicity yields calculated
by West \& Heger (HW, in preparation) with the minimum fallback prescription (see \citealt{c16c}).
The yields of low- and intermediate-mass stars are provided by NuGrid.
We use the IMF of \cite{c03}, with a lower and upper mass boundaries
of 0.1 and 100\,M$_\odot$, and consider that massive stars only contribute to 
the chemical enrichment when their initial mass is between 8 and
30\,M$_\odot$ (see discussion in \citealt{c16b}).
Under these assumptions, $6.80\times10^{-4}$ and $1.48\times10^{-3}$\,M$_\odot$ of
Fe is ejected on average per unit of stellar mass formed when using \blue{K06} and NuGrid yields,
respectively.  For SNe~Ia, we use the yields of \cite{t86} and the delay-time distribution
function described in \cite{c16b}.

In this study, we use the galaxy modeling assumptions described in \cite{c16c} to model the Milky Way.
Figure~\ref{fig_light} shows our numerical predictions for six light elements.  As also shown in
\cite{r10} and in \cite{molla15}, the choice of stellar yields plays a fundamental role in galactic chemical evolution
models.  The breaks seen in the green and black lines at [Fe/H]~$\sim-2.7$, represent the transition between the zero-metallicity
yields of WH (green) and the non-zero metallicity yields of NuGrid (black).
The discontinuities occur because these two sets of yields have not been calculated with
the same code and modeling assumptions, as opposed to \blue{K06} yields, which include both
zero- and non-zero-metallicity stellar models.  The high [Ni/Fe] abundances predicted using NuGrid yields are caused
by an $\alpha$-rich freezout component that is included in the ejecta of some of the CC~SN models.  Such Ni-rich
components are not present in the yields of WH and \blue{K06}, which
explains the significant difference seen in the evolution of [Ni/Fe] when using NuGrid yields (see Figure~\ref{fig_light}). This discrepancy is mostly due to
the stellar remnant mass prescriptions adopted by NuGrid for massive stars. Those prescriptions are designed to reproduce
the neutron star and black hole mass distribution function observed in the Milky Way and therefore generate relatively low remnant
masses for their 12 and 15\,M$_\odot$ stellar models (\citealt{fbw12,p13}).

For each tested set of yields, we tuned
independently mass of gas at early and late times to reproduce the current gas fraction
observed in the Milky Way and to ensure that SNe~Ia appear at [Fe/H]~$\sim-1$ (see \citealt{c16c} for more details).  We
also tuned the strength of galactic outflows to ensure that our predicted metallicity distribution
function (MDFs\footnote{Here, the metallicity distribution function represents the current the number of stars
in the Milky Way having a certain [Fe/H] value in their atmosphere.}) peaks at [Fe/H]~$\sim0$ (see Figure~\ref{fig_mdf}).  Because NuGrid yields
eject more Fe than \blue{K06} yields, stronger outflows are needed with NuGrid yields in
order to provide similar values for the peak of the MDF.  With our model, this results in a flatter age$-$metallicity
relation (see Figure~\ref{fig_age_met}) and thus a narrower MDF (see Figure~\ref{fig_mdf}).

We assume that each NS-NS merger event ejects $10^{-5}$\,M$_\odot$ of Eu, corresponding to a 
total ejected mass of r-process elements of $10^{-2}$\,M$_\odot$ with a Eu mass fraction
of $10^{-3}$ (\citealt{agt07}).  In the next sections, we explore different DTD functions for NS-NS mergers.
As will be shown in Section~\ref{sect_cct}, the total number of NS-NS mergers per stellar mass formed needed to
reproduce the observed Eu abundances in the Milky Way is $5.01\times10^{-5}$ when using \blue{K06} yields, and
$7.24\times10^{-5}$ when using NuGrid yields (see Section~\ref{sect_psm} for uncertainties).  For our study, C16, the circles and triangles shown in Figure~\ref{fig_EuFe}
present our results using NuGrid and \blue{K06} yields, respectively.

\subsection{Normalized Comparison}
\label{sect_nc}
As shown in the previous sections, different GCE models
and simulations generally use different IMFs with different mass boundaries, different
stellar yields, different progenitors for massive stars and NS-NS mergers, and different DTD functions and normalizations
for the NS-NS merger rates.  To provide a relevant comparison between the considered studies,
we have corrected $N_\mathrm{NS-NS}$, the derived number of NS-NS mergers per unit of stellar mass formed,
in order to normalize the studies to the same Eu and Fe yields.  For each study, the normalized value, $N'_\mathrm{NS-NS}$,
is given by
\begin{equation}
\label{eq_norm}
N'_\mathrm{NS-NS} = N_\mathrm{NS-NS}\left(\frac{M_\mathrm{Eu}}{M'_\mathrm{Eu}}\right)\left(\frac{M'_\mathrm{Fe}}{M_\mathrm{Fe}}\right),
\end{equation}
where $M_\mathrm{Fe}$ and $M_\mathrm{Eu}$ are the masses of Fe and Eu ejected by massive stars
and NS-NS mergers, respectively, originally adopted in the study (see Figure~\ref{fig_EuFe}).  $M'_\mathrm{Fe}$
and $M'_\mathrm{Eu}$ represent the normalized yields common to all studies.  Here we chose $M'_\mathrm{Fe}=3.35\times10^{-4}$\,M$_\odot$,
the median value of the $M_\mathrm{Fe}$ distribution, and $M'_\mathrm{Eu}=10^{-5}$\,M$_\odot$, representing
a Eu yield of $10^{-3}$ with a total mass of r-process ejecta of $10^{-2}$\,M$_\odot$.  We refer
to Section~\ref{sect_cwpsm} for the error bars of our normalization.

\begin{figure}
\begin{center}
\includegraphics[width=3.25in]{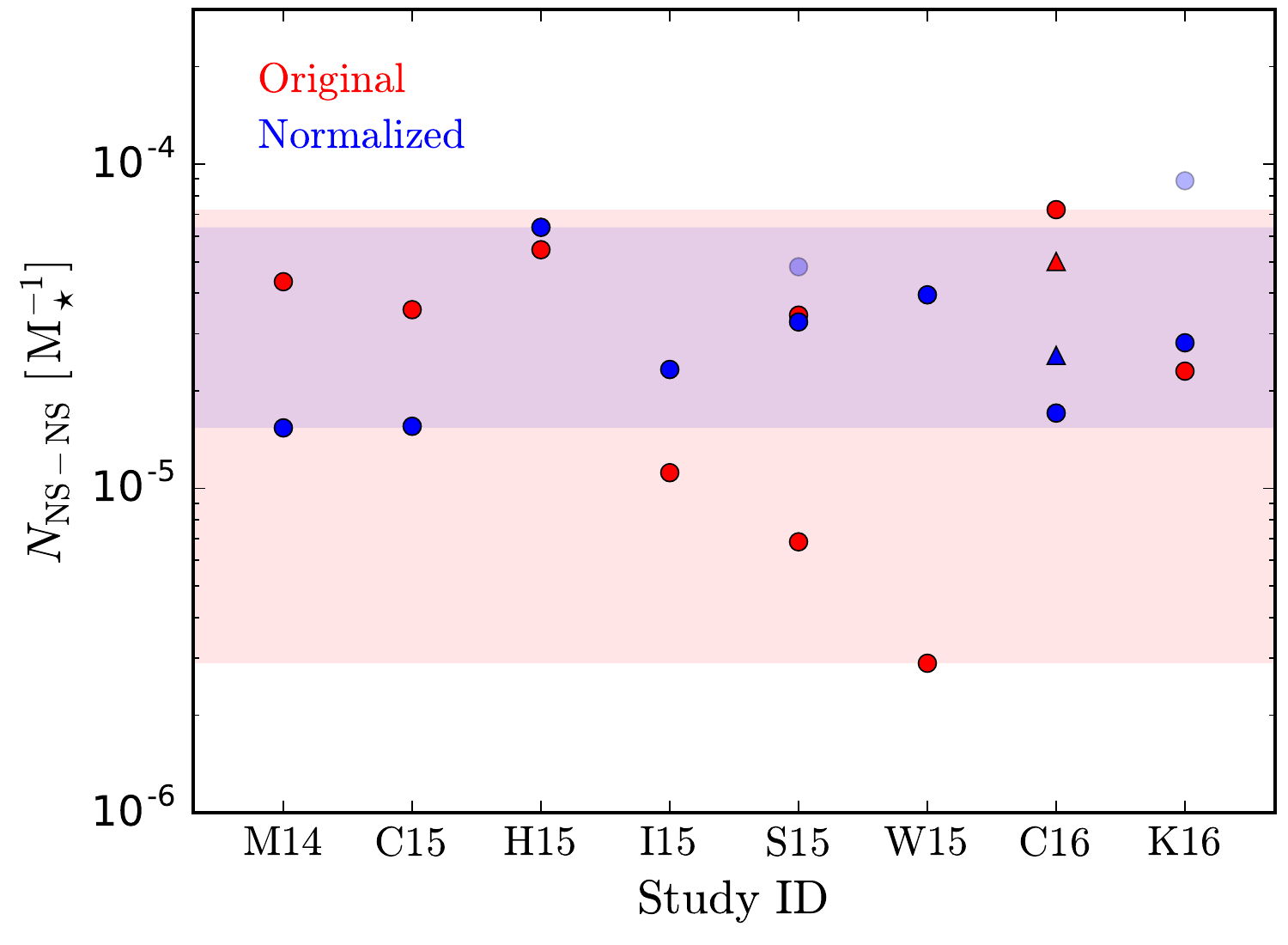}
\caption{Number of NS-NS mergers, per unit of stellar mass formed, needed in different chemical evolution studies in order to reproduce the Eu abundances in the Milky Way.
The red circles represent the original values extracted from each study while the blue circles show the normalized values as if all studies used the
same mass of Eu ($10^{-5}$\,M$_\odot$) ejected per NS-NS merger event and the same mass of Fe ($3.35\times10^{-4}$\,M$_\odot$) ejected by
massive stars per stellar mass formed.  The red and blue shaded areas highlight the range of values given by the considered studies.
\blue{S15} and \blue{K16} originally showed a final Eu abundances that was larger than the solar [Eu/Fe] composition.
To be consistent with other studies, a correction factor has thus been applied to scale their final [Eu/Fe] ratios down to zero.  The transparent blue circles represent the normalized values before the correction.}
\label{fig_N_NSM}
\end{center}
\end{figure}

This normalization is not complete, as it does not include corrections for different implementations of SNe~Ia,
which are significant contributors of Fe at [Fe/H] $\gtrsim-1$.  Nevertheless, as shown in Figure~\ref{fig_N_NSM},
this normalization significantly reduces the level of scatter seen between the various studies, from a factor of $\sim25$
(red symbols) down to a factor of $\sim4$ (blue symbols).  An additional correction has been applied to
\blue{S15} and \blue{K16} since their [Eu/Fe] value at [Fe/H] $=0$ was originally greater than zero.  The normalized values,
before the application of this additional correction, are shown as transparent blue circles.  The range shown in blue in
Figure~\ref{fig_N_NSM} overlaps with the values derived from Eu abundances in ultra-faint dwarf galaxies (\citealt{bhkp16}).

Our work, C16, originally showed the highest value for $N_\mathrm{NS-NS}$ because we used the highest Fe yields
(see lower panel of Figure~\ref{fig_EuFe}).  Therefore, compared to other studies, we needed more NS-NS mergers to
counterbalance the Fe ejecta from massive stars.  However, if we had used the normalized $M'_\mathrm{Fe}$, which is
lower than the original $M_\mathrm{Fe}$, less NS-NS mergers would have been necessary to reach the same Eu abundances.
\blue{W15} originally showed the lowest value for $N_\mathrm{NS-NS}$ because they used one of the highest Eu yields
(see middle panel of Figure~\ref{fig_EuFe}) and one of the lowest Fe yields (see lower panel of Figure~\ref{fig_EuFe}).
This means \blue{W15} did not require as many NS-NS mergers as in other studies to each the same [Eu/Fe] ratios.  By using lower
Eu yields and larger Fe yields, ($M_\mathrm{Eu}/M'_\mathrm{Eu}$) and ($M'_\mathrm{Fe}/M_\mathrm{Fe}$)
both become larger than 1, which increases the number of NS-NS mergers needed to reproduce the Eu abundances
(see equation~\ref{eq_norm}).

We refer to Section~\ref{sect_disc} for a discussion on different sources of uncertainties that can affect the normalization process.

\section{Analytical Delay-Time Distributions}
\label{sect_DTD}
In this section, we use {\tt OMEGA} and explore different DTD functions commonly used
for NS-NS mergers in GCE studies to investigate how they affect our ability to reproduce the global chemical evolution trend of Eu
in the Milky Way.  We recall that we focus on Eu in order to compare our results with the previous
studies described in Section~\ref{sect_GCE}.

\subsection{Constant Coalescence Timescale}
\label{sect_cct}
\begin{figure}
\begin{center}
\includegraphics[width=3.25in]{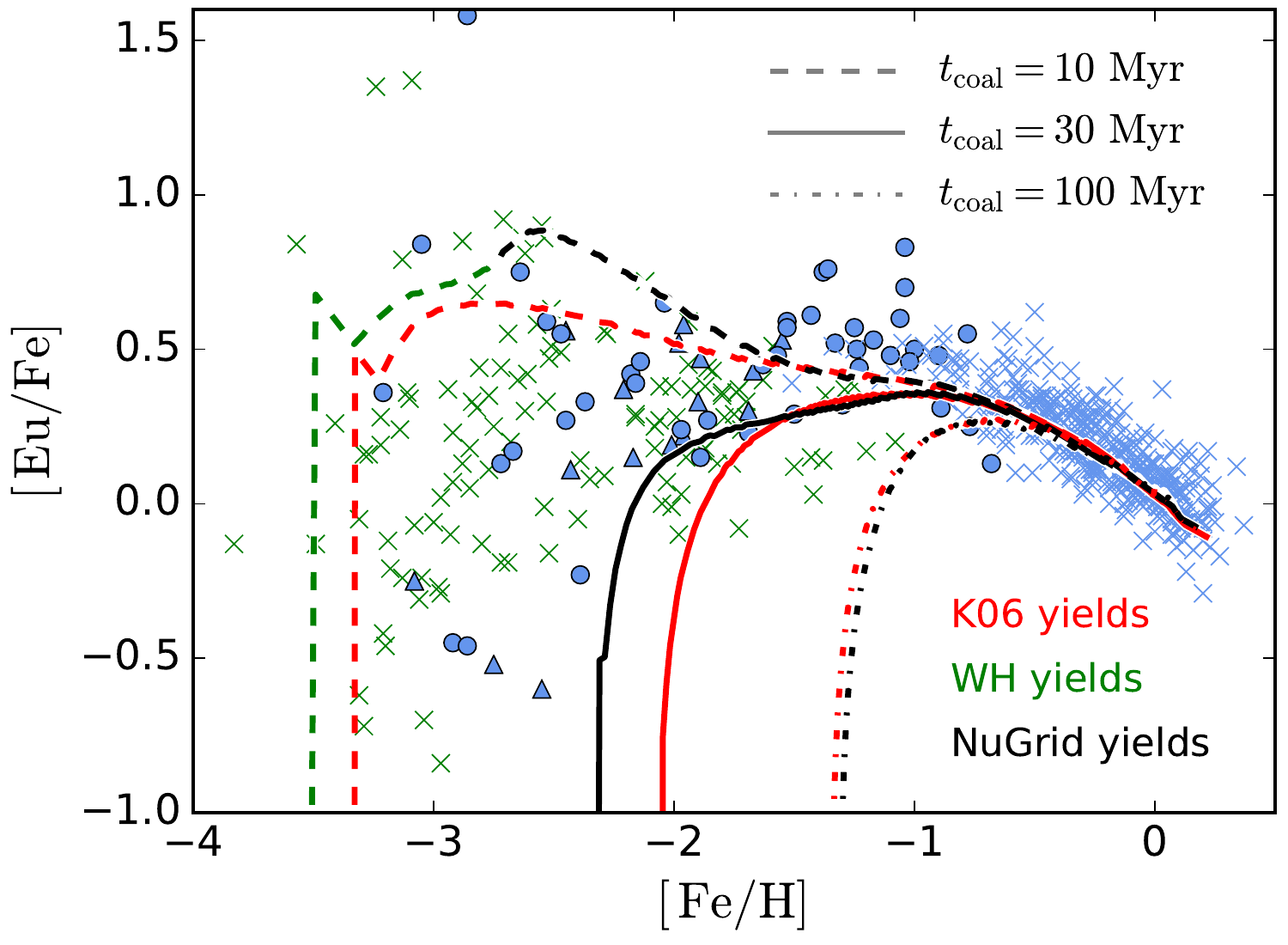}
\caption{Predicted chemical evolution of Eu, for different constant coalescence timescales for NS-NS mergers (different line styles), using the yields calculated by \protect\citet[red lines]{k06}, West \& Heger (in preparation, green lines), and NuGrid (black lines).  In this study, the constant coalescence timescale represents the delay between star formation and the appearance of all NS-NS mergers in a stellar population.  Observational data come from \protect\citet[blue triangles]{r09}, \protect\citet[blue circles]{h12} \protect\citet[green crosses]{r14}, and \protect\citet[blue crosses]{bb16}.}
\label{fig_t_coal}
\end{center}
\end{figure}

Figure~\ref{fig_t_coal} shows our numerical predictions using constant coalescence timescales since star formation, $t_\mathrm{coal}$, after which
all NS-NS mergers in a stellar population release their ejecta.  As found by almost all previous studies, $t_\mathrm{coal}$ has an impact on when,
or at which [Fe/H], NS-NS mergers first appear.  The shape of our numerical predictions are similar to
one generated by other models that either use a constant coalescence timescale (e.g., \citealt{a04}; \blue{W15}) or assume
that all NS-NS mergers occur within a very short period of time (e.g., \blue{M14}; \blue{C15}).  A bend in the evolution of [Eu/Fe] 
at [Fe/H]~$\sim-1$ is clearly visible and originates from the iron ejected by SNe~Ia.
Furthermore, our results are consistent with the ones of \blue{M14} (see their Figure~2) in the sense that the coalescence 
timescale parameter can shift the appearance of NS-NS mergers on the [Fe/H] axis, but cannot change the
[Eu/Fe] ratio at later times (see Figure~\ref{fig_t_coal}).  This conclusion, however, is only valid for relative short coalescence timescales.
In Figure~\ref{fig_t_coal}, the total number of NS-NS mergers
per unit of stellar mass formed was $5.01\times10^{-5}$ when using \blue{K06} yields and
$7.24\times10^{-5}$ when using NuGrid yields, for all chosen values for the
coalescence timescale. All of the models cited in this paragraph, including ours, 
do not include hydrodynamics.  We note, however, that \blue{H15} used constant 
coalescence timescales in their hydrodynamic simulations.

When $t_\mathrm{coal}=100$\,Myr (dot-dashed lines in Figure~\ref{fig_t_coal}), our NS-NS mergers appear at a [Fe/H] value that is similar to
one found in \blue{M14}.  However, when $t_\mathrm{coal}$ is shorter, our NS-NS mergers tend to appear at lower [Fe/H] compared to \blue{M14}.
This can be caused by the fact that in \blue{M14}, the coalescence timescale represents the time span between the CC~SNe and NS-NS mergers events,
whereas in our model it represents the time span between star formation and NS-NS mergers events.
Alternatively, it can be caused by different gas and star formation efficiency treatments at early time.  As a matter of fact,
because of the abrupt slope of the age$-$metallicity relation when [Fe/H] is below $-1$ (see Figure~\ref{fig_age_met}), a slight delay in the enrichment process, or 
a slight modification of the enrichment efficiency, can induce a major change on when NS-NS mergers first appear on the [Fe/H] axis.

Many uncertainties are associated with the first Gyr of evolution in our model (e.g., very low-metallicity yields, star formation efficiency,
gas content and metal mixing, and strength of galactic inflows and outflows).  Furthermore, we do not think that our one-zone model
is suited to reproduce the evolution of our Galaxy at early time.  Stochastic processes and a proper treatment of how galaxies 
assemble in the early Universe through galaxy mergers is currently not included in {\tt OMEGA} (but see our recent developments 
in \citealt{c16d}).  For this reason, we conclude that it is more reliable to use the total number of NS-NS mergers, 
rather than when they appear, to investigate whether or not NS-NS mergers can be the dominant source of r-process elements.  This 
number can directly be compared with population synthesis models (see Section~\ref{sect_psm}).  In addition, as will be shown in Section~\ref{sect_predcmr},
the predicted cosmic NS-NS merger rate density, which can be compared with upcoming Advanced LIGO measurements, is more sensitive to the total number of merger events than to the
choice of the DTD function.

\subsection{Power Law Distribution}
\label{sect_pldtd}
\begin{figure}
\begin{center}
\includegraphics[width=3.25in]{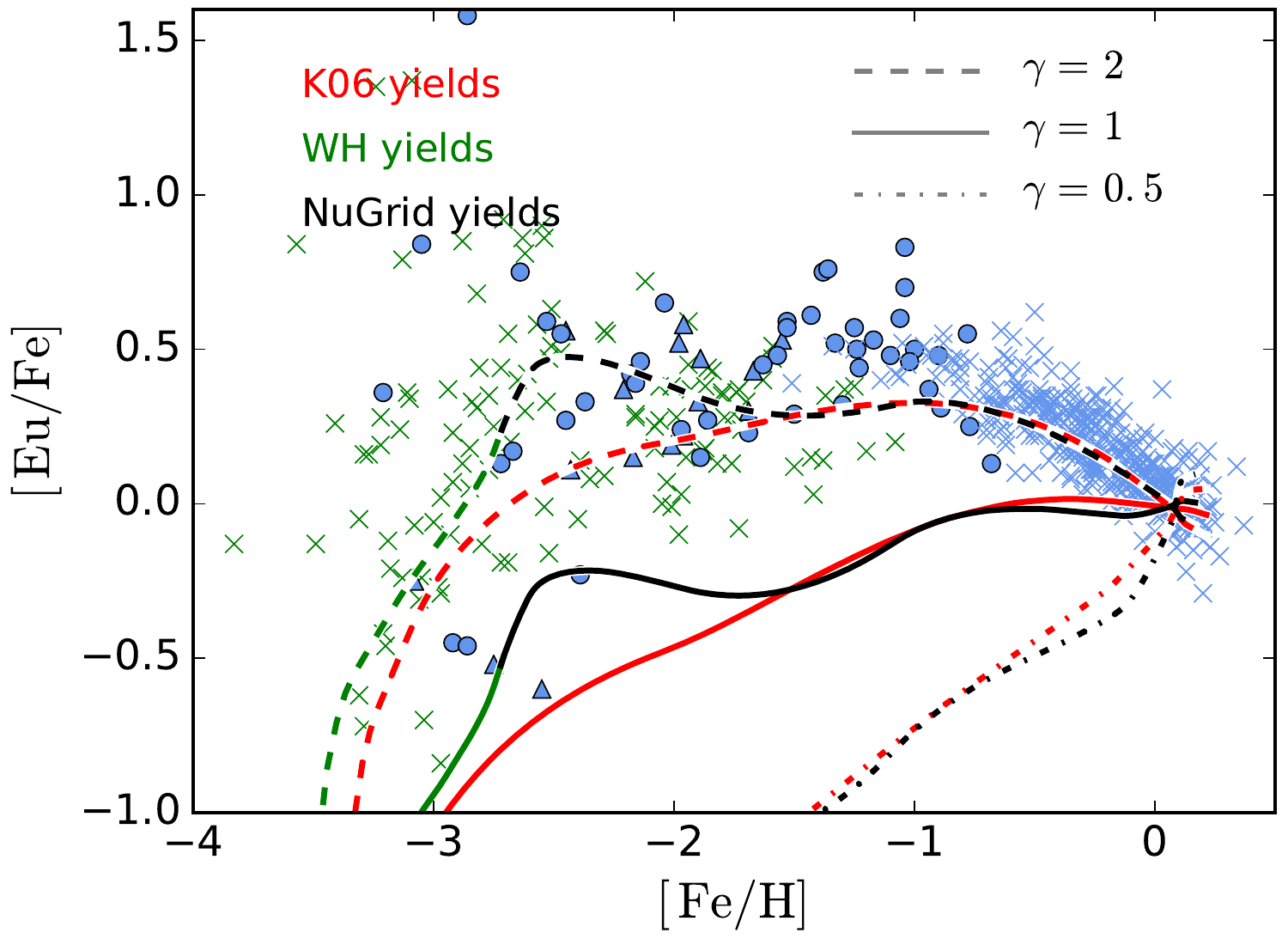}
\caption{Same as Figure~\ref{fig_t_coal}, but using a power-law delay-time distribution function for NS-NS mergers in the form of $t^{-\gamma}$, applied from 10\,Myr to 10\,Gyr, with different values for $\gamma$ (different line styles).}
\label{fig_power_law}
\end{center}
\end{figure}

A more realistic implementation is to assume that NS-NS mergers in a stellar population are distributed within
several Gyr instead of occurring all after one fixed, constant coalescence time.  Here we explore the delay-time
distribution (DTD) function in the form of $t^{-\gamma}$ with a lower and upper time boundaries
of 10\,Myr and 10\,Gyr.  We refer to Section~\ref{sect_psm} for different
forms.  Figure~\ref{fig_power_law} shows our numerical predictions using different values for the power-law
index, $\gamma$.  In all cases, we normalized the DTD functions to generate the same number of NS-NS mergers,
per unit of stellar mass formed, as in Section~\ref{sect_cct}.  As shown in the last figure, our
one-zone model cannot reproduce the Eu trend when $\gamma=1$ (solid lines), which is, however,
a value motivated by the work of \cite{d12}.  Our predictions are worst when $\gamma=0.5$ (dot-dashed lines),
since NS-NS mergers are further delayed compared to the case where $\gamma=1$.  We reasonably reproduce observations 
when $\gamma=2.0$ (dashed lines), since this abruptly decreasing power-law brings our implementation closer to the
constant coalescence timescale assumption (see Figure~\ref{fig_t_coal}).

The small variations seen in Figure~\ref{fig_power_law} at [Fe/H]~$>0$ are caused by
different amount of Eu lost by galactic outflows, which are more powerful
at the beginning of our simulations.  More Eu is thus kept in the
system when NS-NS mergers mainly occur at later times (dot-dashed lines).

\subsubsection{Non-Uniform Mixing}
As opposed to {\tt OMEGA}, cosmological zoom-in hydrodynamic
simulations (\blue{S15}, \citealt{vv15}) and semi-analytical models of galaxy formation within
a cosmological context (\citealt{ks16}) can reproduce metal-poor stars, at least
down to [Fe/H]\,$\sim-3$, with $\gamma=1$.  This highlights major differences
between one-zone models and more sophisticated simulations.  In particular, 
hydrodynamic simulations self-consistently follow the non-uniform mixing of
metals at early time, which generates significant scatter in the age$-$metallicity relation
(see also \citealt{k11}; \blue{H15}).  \blue{S15} compared their original results with an homogeneous-mixing
version of their simulations, which is similar to our one-zone model.  They found
that Eu can only be recycled in very-low metallicity stars when stellar and NS-NS merger ejecta are
not uniformly mixed within the galactic gas (see their Figure~7).  However, non-uniform
mixing is not the only important ingredient to consider in order to address the early evolution
of our Galaxy.  Inhomogeneous chemical evolution models, such as the
ones of \blue{C15} and \blue{W15}, still have difficulties to reproduce the Eu abundances at [Fe/H]~$\lesssim-2$
with reasonable coalescence timescales when considering NS-NS merger as the exclusive
r-process site.

\subsubsection{Cosmological Context}
Another reason why our one-zone model fails to reproduce the Eu trend 
at early time, when using a power law for the DTD function for NS-NS mergers with $\gamma=1$ and 0.5, is because
we ignore the hierarchical nature of how galaxies form in a
cosmological context.  One-zone models consider a single gas reservoir
associated with one galaxy while in reality, massive galaxies such as the Milky Way
are the results of many galaxy mergers (e.g., \citealt{v14,schaye15,g16}). 
Within this framework, the early evolution of the Milky Way should therefore be represented by many 
low-mass progenitor galaxies.

Recently, \blue{H15} shown with hydrodynamic simulations
that NS-NS mergers can enrich stars at [Fe/H]~$\lesssim-3$ in dwarf galaxies when
the star formation efficiency is lowered compared to more massive systems,
which slows down the evolution of [Fe/H] at early time (see also \blue{I15}).
Although \blue{H15} assumed a constant coalescence timescale instead of a power law for
the DTD function of NS-NS mergers, they were able to reproduce metal-poor stars even when no NS-NS merger
occurred before 100\,Myr (see their Figure~7).  This is not possible with simpler models, as
shown by the dot-dashed lines in our Figure~\ref{fig_t_coal} (see also \citealt{a04}; \blue{M14}; \blue{C15}; \blue{W15}).
\cite{ks16} also demonstrated, using a semi-analytical model within a cosmological context, that Ba and Eu
are better reproduced with NS-NS mergers if the star formation efficiency decreases with the mass of
progenitor galaxies. 

\subsubsection{Shape of Numerical Predictions}
As shown in Figure~\ref{fig_t_coal} and \ref{fig_power_law}, the shape of
numerical predictions are different when assuming a constant coalescence
timescale or a power law for the distribution of NS-NS mergers as a function of time.
When using a power law with an index of $-1$ or $-0.5$, it is difficult to 
recover the knee observed at [Fe/H]~$\sim-1$ in the evolution of Eu.  
Even if our GCE model is relatively simple, this lack
of decreasing trend above this metallicity is also seen in the hydrodynamic simulations of \blue{S15}.
The knee is visible in the simulations of \citealt{vv15}, but only at [Fe/H]~$\sim0$.
This difficulty is also reported by \cite{ks16}.  On the other hand, the knee can
easily be reproduced when using coalescence timescales after which all NS-NS mergers
occur in a stellar population (see Figure~\ref{fig_t_coal} and \blue{M14}; \blue{C15}; \blue{W15}).
If NS-NS mergers are actually the main source of r-process elements, 
this suggests that NS-NS mergers need to occur on short timescales.

\subsection{Constraints for Neutron Star Mergers}
As discussed in the previous sections, it seems like non-uniform mixing and a proper treatment of how galaxies 
assemble in the early Universe are essential ingredients to reproduce
low-metallicity stars and to understand how NS-NS merger ejecta can be recycled
in the early stage of our Galaxy.  However, at higher [Fe/H], GCE simulations in general are
less sensitive to those aspects and simple models like {\tt OMEGA} can provide
valuable insights (see also Figure~6 in \blue{S15}).  Although our one-zone model
is not suited to reproduce metal-poor stars, our predictions always end
roughly at [Eu/Fe]~$\sim0$, regardless of the choice of implementation
for the delay of NS-NS mergers (see Figures~\ref{fig_t_coal} and \ref{fig_power_law}).
To constrain NS-NS mergers and investigate whether or not they can be the
dominant source of r-process elements, we believe that the total number
of neutron star mergers needed to ultimately reach [Eu/Fe]~$\sim0$ 
at [Fe/H]~$\sim0$ is a more robust and universal constraint than evaluating the capacity
of models to reproduce metal-poor stars, as it is less affected by the
various sources of uncertainty associated with the early evolution of
our Galaxy.

\section{Population Synthesis Models}
\label{sect_psm}
Here we introduce our population synthesis models, present their predicted compact binary merger rates, and discuss their impact on GCE.

\subsection{Code Description}
\label{model}

Population synthesis calculations were performed using {\tt StarTrack}  
\citep{Belczynski2002,Belczynski2008a}.  This code has recently been
improved for the evolution of massive stars.  It now includes better assumptions for the
evolution of the common envelope (CE) phase \citep{d12}, the masses of compact objects
produced by CC~SNe \citep{fbw12,bwf12}, the initial conditions for stellar binaries
(constrained by observations, \citealt{deMink2015}), and for the evolution of star
formation and the average metallicity in the Universe (also constrained by observations, \citealt{Belczynski2016b}).
This last improvement is used in Section~\ref{sect_cmr} to predict the cosmic
compact binary merger rate densities.

Our calculations are originally based on analytic fits made to the non-rotating stellar models presented in
\citet{Hurley2000}. However, we updated these models with revised 
stellar wind prescriptions \citep{Vink2011} and a new compact-object formation 
scheme \citep{fbw12}.  We have begun calibrating our evolution with 
calculations performed with modern stellar models \citep{Pavlovskii2016}.

\subsubsection{Initial Setup}
All massive stars with a zero age main sequence mass ($M_{\rm ZAMS}$) greater than $7$\,--\,10\,M$_\odot$
are assumed to be the progenitors of neutron stars and black holes.
The initial parameters for massive stars in binary systems are guided by recent
observations of O/B binaries \citep{Sana2012,Kobulnicky2014}. The mass of the primary stars is 
chosen from a three-component broken power-law initial mass function with an index of $-2.3$
for massive stars. A flat mass 
ratio distribution is used to calculate the mass of the secondary stars.  Binaries are assumed 
to form predominantly on close and nearly circular orbits. We assume that the binary fraction is
$100\%$ for stars more massive than 10\,M$_\odot$ and $50\%$ for less massive stars.

As mentioned above, the stellar models used in our study do not include the effects of
rotation on their evolution. However, we include the impact of rotation with
estimates of the tidal interactions between the stars and their binary orbit. 
We assume that our stars rotate at moderate velocities ($200$\,--\,$300$ km s$^{-1}$).
We do not consider the small fraction of massive stars that may rotate at very high speeds ($\sim 600$ km s$^{-1}$). 
For such rapidly rotating stars, the effects of rotation on their evolution
need to be included in evolutionary calculations 
\citep{Marchant2016,deMink2016,Eldridge2016,Woosley2016}.

\subsubsection{Binary Evolution}
Stellar spins (and thus binary orbits) are affected by magnetic
braking when stars have significant convective envelopes. Additionally, we
account for the orbital changes due to the mass lost by stellar winds and angular
momentum loss due to the emission of gravitational waves (important only for very compact binaries). 
The development and dynamical stability of the Roche lobe overflow
(RLOF) is judged based on the following: 
binary mass ratio, evolutionary stage of the donor, response to mass loss, and 
behaviour of the orbital separation in response to mass transfer and angular 
momentum loss~\citep{Belczynski2008a}.

During stable RLOF, we assume that half of the mass is accreted onto the companion, 
while the other half is lost from the binary with the specific angular momentum (we adopt the rather effective angular momentum loss with $j_{\rm loss}=1.0$ defined in
~\citealt{Podsiadlowski1992}). Unstable mass transfer is assumed to lead 
to a CE. This is treated using the energy balance formalism with an effective 
conversion of orbital energy into envelope ejection ($\alpha=1.0$).  The 
envelope binding energy depends on the mass, radius, and metallicity of the donor star.
During the CE phase, neutron stars 
and black holes accrete at $10\%$ of the Bondi-Hoyle rate~\citep{Ricker2008,Macleod2015}.

We account for mass loss, neutrino loss, 
and natal kicks during SNe. These explosions affect the binary orbits and can,
for specific configurations, unbind and disrupt binary systems.

\subsubsection{Neutron Stars and Black Holes}
\label{sect_ne_bl_pop_syn}
The mass of each compact object is based on a selection of hydrodynamical SN 
models that are initiated from detailed stellar evolutionary 
calculations~\citep{Woosley2002,heger03,Fryer2006,Limongi2006,Dessart2007,
Poelarends2008,Young2009}. The resulting mass of a compact object is
based on the stellar mass at the time of core collapse and on the final mass of its 
carbon-oxygen core~\citep{fbw12}. In this study, we use a rapid SN explosion prescription that 
reproduces the observed mass gap between neutron stars and black holes \citep{bwf12}. Neutron stars are formed with
masses in the range of $1.1$\,--\,$2.5$\,M$_\odot$, while black holes form with masses in the range of
$5$\,--\,$94$\,M$_\odot$.  The upper limit for the mass of a black hole is set by the strength of stellar
winds and their dependence on metallicity~\citep{Belczynski2010b}.

For single stars, with our formulation, there is a strict delimitation between
neutron stars and black holes based on the ZAMS mass of the compact object progenitors. 
This mass threshold depends on metallicity and is found at
$M_{\rm ZAMS}\approx20$\,M$_\odot$ for solar metallicity ($Z=0.02$) progenitors.
We note that binary interactions may significantly affect this limit. 
As a matter of fact, stars as massive as $100$\,M$_\odot$ can form either neutron stars 
or black holes depending on their specific binary configurations
\citep{Belczynski2008c}. In other words, our simulations include non-monotonic 
formation of neutron stars and black holes with respect to progenitor initial mass, with neutron stars and black holes
mixed up in a wide range of initial masses ($M_{\rm ZAMS}\approx10$\,--\,$100$\,M$_\odot$).
This prescription has been designed to reproduce observations of compact objects.

\subsubsection{Selection of Models}
\label{sect_sel_models}
In this study, we use three models that were calculated and presented in
\cite{Belczynski2016b}, which are the standard model (M1), the optimistic model (M2), and a model
with low natal kicks (M6). The standard model includes our best choices on the 
various parts of uncertain evolutionary physics as described above. 
In particular, the natal kicks for neutron stars and black holes are adopted from the velocity 
distribution measured for single pulsars in our Galaxy \citep{Hobbs2005} -- the 
natal kicks are drawn from a $1$-D Maxwell distribution with dispersion of $\sigma=265$\,km\,s$^{-1}$ 
and an average $3$-D speed of $\sim 400$\,km\,s$^{-1}$. However, we lower the natal kick 
value according to the amount of fallback calculated for each compact object formation. 
For neutron stars, the amount of fallback is rather small and they receive high natal
kicks. For very massive black holes ($\gtrsim 10$\,M$_\odot$), all or almost all material
falls back onto the black holes and thus receive no natal kick. For moderate mass
black holes, the natal kick value is also moderate \citep{fbw12}.

In the standard model M1, we assume that stars that evolve through the Hertzsprung gap, mostly 
with radiative envelope, cannot survive a CE phase and merge before becoming double compact binaries~\citep{Belczynski2007}.
The two alternative models differ from the standard M1 model by only one aspect.
In model M2, we allow Hertzsprung gap stars to initiate and a CE and 
to survive it. This effectively enhances the formation of all types of double
compact object mergers~\citep{Belczynski2007,Belczynski2016b}.
In model M6, all compact objects, both neutron stars and black holes, receive a 
low natal kicks with $\sigma=70$\,km\,s$^{-1}$. We do not modify natal kicks with the
fallback factor. This means that neutron stars receive on average smaller natal kicks while
massive black holes receive larger natal kicks, compared to model M1. 

\begin{figure*}
\begin{center}
\includegraphics[width=7in]{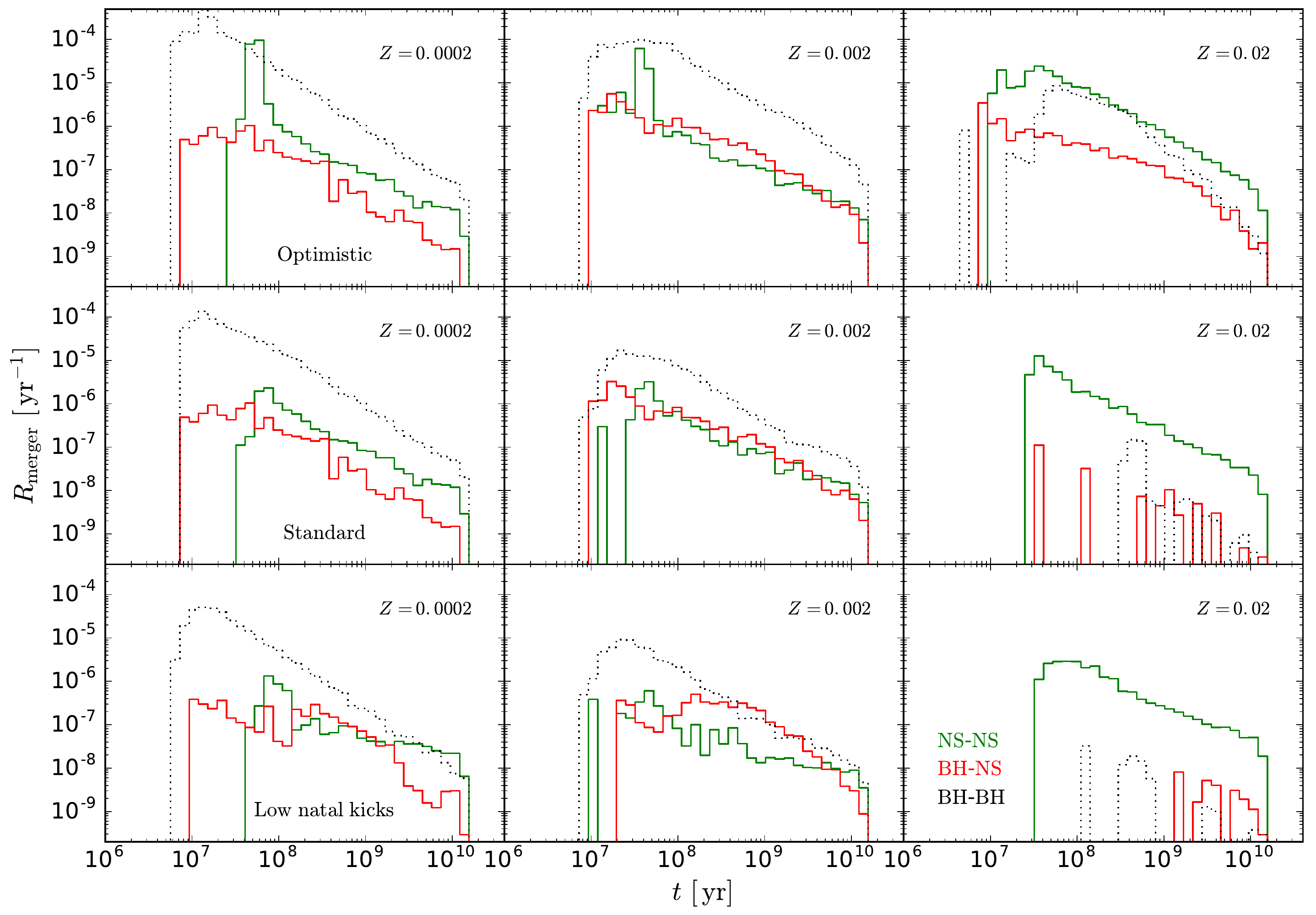}
\caption{NS-NS (green), BH-NS (red), and BH-BH (black) merger rates, as a function of time, predicted by population synthesis models for a stellar population of $2.8\times10^8$\,M$_\odot$.  Different columns represent different metallicities, while different rows represent models with different evolutionary assumptions.}
\label{fig_merger_rate}
\end{center}
\end{figure*}

Recent detailed evolutionary calculations seem to indicate that Hertzsprung gap
stars should lead to a stable RLOF rather than to a CE \citep{Pavlovskii2016}.
This scenario would thus lead to a larger number of non-merging
compact objects compared to the M2 model. This is why
we call the M2 model optimistic. Actually, in cases of very massive stars leading
to black hole - black hole mergers, this model already appears be excluded by 
advanced LIGO measurements~\citep{Belczynski2016b}. However, we cannot
exclude yet the possibility that this M2 model is consistent for lower mass stars (neutron star
progenitors), since our predicted cosmic NS-NS merger rate density for this model is
$\sim 250$ Gpc$^{-3}$ yr$^{-1}$ and is currently below the upper limits
expected by advanced LIGO (see Section~\ref{sect_cmr}).

A more detailed study of extreme cases in population synthesis models
will soon be conducted.  Although our selection of models provides a
view of the evolutionary uncertainties associated with the models described in \cite{Belczynski2016b},
we warn that they are not representative of the entire range of plausible
scenarios.

\subsection{Predicted Merger Rates}
Figure~\ref{fig_merger_rate} shows the predicted merger rates using our optimistic (upper panels),
standard (middle panels), and low natal kick (lower panels) models, for a stellar population
with a total mass of $2.8\times10^8$\,M$_\odot$.  Although only three metallicities are presented
in this last figure, the NS-NS (green lines), BH-NS (red lines), and BH-BH (black lines) merger rates
have been calculated for 32 metallicities.  At low metallicity, BH-NS and BH-BH mergers 
appear earlier than NS-NS mergers.  This is because black holes typically have a larger mass
at lower metallicities (\citealt{Belczynski2010b}), which leads to shorter merger times compared
lighter-mass NS-NS systems (\citealt{peters64}).

Figure~\ref{fig_nb_per_m} illustrates the total number of
mergers per unit of stellar mass formed, which represents the normalization of the merger rates
presented in Figure~\ref{fig_merger_rate}.  Above $Z\sim10^{-2}$, there is an increase in 
the number of NS-NS mergers and a drop in the number of BH-NS and BH-BH mergers.  This
is caused by a higher relative frequency of black holes at low metallicity, and a higher
relative frequency of neutron stars at high metallicity.

\subsection{Comparison with GCE Studies}
\begin{figure}
\begin{center}
\includegraphics[width=3.25in]{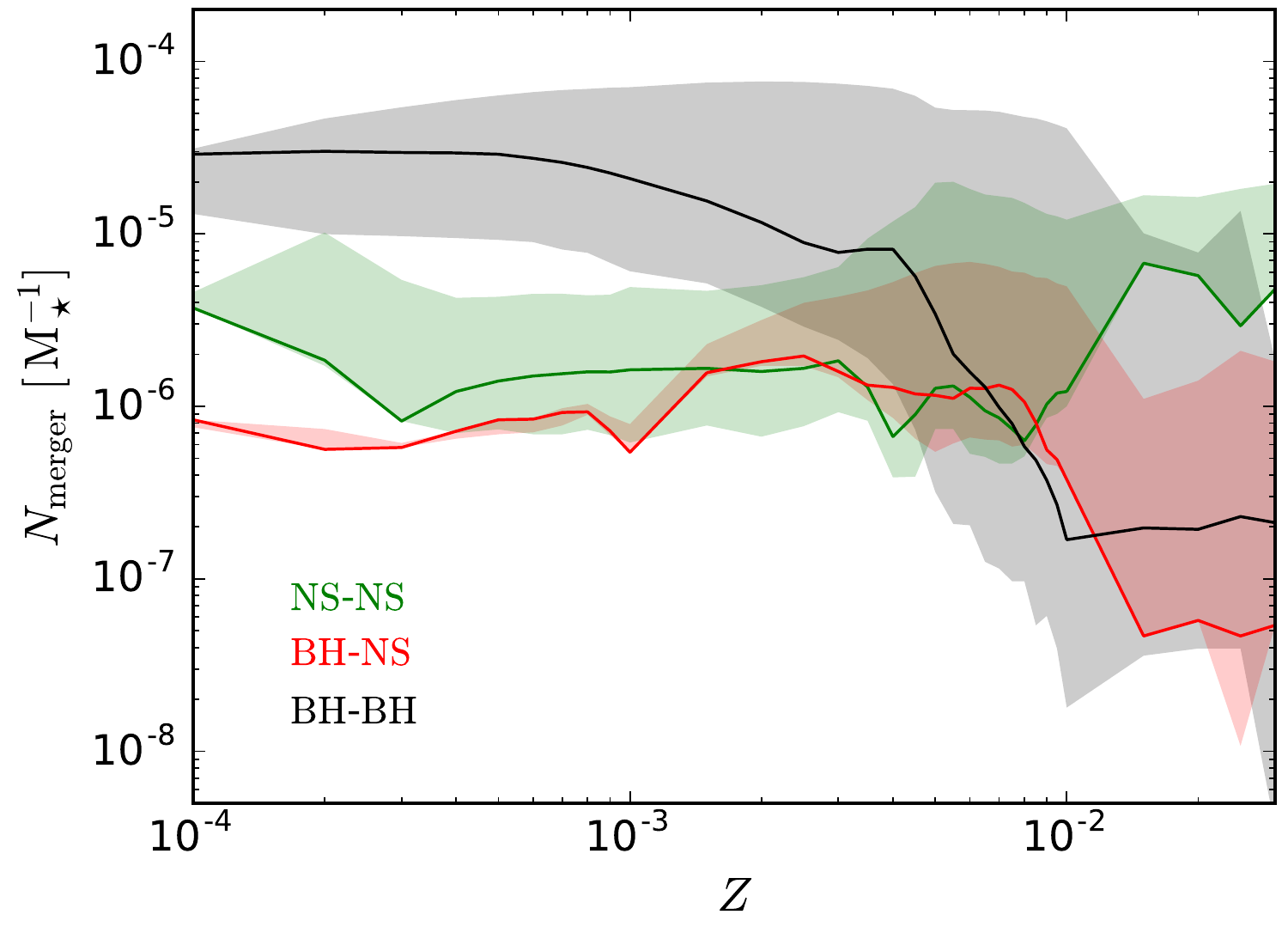}
\caption{Number of NS-NS (green), BH-NS (red), and BH-BH (black) merger, per unit of stellar mass formed as a function of metallicity, predicted by population synthesis models.  The solid lines represent the standard model while the shaded areas show the range of possible predictions defined by the two alternative models.}
\label{fig_nb_per_m}
\end{center}
\end{figure}

In Figure~\ref{fig_N_NSM_pop_GCE}, we compare the normalized numbers of
NS-NS mergers extracted from GCE studies (blue) with the numbers predicted by population synthesis 
models (green).  The blue circles have been computed assuming that each NS-NS 
merger ejects $10^{-5}$\,M$_\odot$ of Eu and that CC~SNe eject on average
$3.35\times10^{-4}$\,M$_\odot$ of Fe per units of stellar mass formed (which is the
median value taken from the lower panel of Figure~\ref{fig_EuFe}).  To consider
the wide range of Eu and Fe yields and to highlight the possible
numbers of NS-NS mergers that can reproduce the same [Eu/Fe] level
as the blue circles, we added in Figure~\ref{fig_N_NSM_pop_GCE} in the blue vertical lines.
To do so, we considered the lowest and largest Fe
yields presented in Figure~\ref{fig_EuFe}, assumed that the total mass ejected by NS-NS mergers
ranges between 0.001 and 0.025\,M$_\odot$ (see Section~\ref{sect_me}), and used equation~(\ref{eq_norm}) to calculate
the lower and upper limits.  We did not account for the uncertainties in the choice of solar normalization
and in the mass fraction of Eu in the solar r-process residual, as they are not significant compared
to the uncertainty associated with the total mass ejected by NS-NS mergers.  The dark green shaded area
represents the range of values predicted by the standard population synthesis model at different metallicities,
while the light green shaded area also includes our two alternative models (see Figure~\ref{fig_nb_per_m}).

As seen in Figure~\ref{fig_N_NSM_pop_GCE}, all GCE studies overlap with the optimistic
population synthesis model, and half of them overlap with the standard model as well.
But in general, it appears that GCE systematically needs more NS-NS mergers than
what is predicted by population synthesis models.  We warn that the comparison shown
in Figure~\ref{fig_N_NSM_pop_GCE}  can be misleading, as $N_\mathrm{NS-NS}$ is
constant in GCE studies while it is metallicity-dependent in population synthesis models.
 
\subsection{Combining OMEGA and Population Synthesis}
\label{sect_cwpsm}
To better capture the impact of the metallicity-dependent predictions of
population synthesis models in a GCE context, we included them in {\tt OMEGA}
as an input parameter.  We note that \cite{mv14,mv16} also introduced population
synthesis predictions in a GCE code.  Figure~\ref{fig_gce_pop} shows our GCE predictions using
NS-NS mergers only (upper panel), BH-NS mergers only (middle panel), and both NS-NS and BH-NS mergers
(lower panel).  In all cases, the normalization of the DTD functions, or the total number of 
compact binary mergers, is directly taken from population synthesis predictions.
To better visualize the role of compact binary mergers, we switched off
the contribution of AGB stars in the evolution of Eu, as they provided a 
minimum floor value of [Eu/Fe]~$\sim-1$ at high [Fe/H] and prevented us
from visualizing the contribution NS-NS and BH-NS mergers.

The solid lines represent the standard model and the dark shaded areas show the range
of predictions generated by assuming different ejected masses for the merger events.
Alternative models are treated in the same manner and are shown with the dashed lines
and the light shaded areas.  We assumed that each BH-NS merger
event ejects between 10$^{-4}$ and 10$^{-1}$\,M$_\odot$ of r-process material (see Section~\ref{sect_me}).
We then assumed, as a fiducial case, that only 20\,\% of all BH-NS mergers are able to eject material.
However, we allowed this pourcentage to vary from 2\,\% to 40\,\%, representing low and high black hole spins (see Section~\ref{sect_ej_BH_NSMs}).
This extra source of variation has been combined with
the range of total masses ejected by BH-NS mergers in order to generate the shaded areas in Figure~\ref{fig_gce_pop}.

BH-NS mergers appear earlier than NS-NS mergers at low metallicity, which was also
seen in Figure~\ref{fig_merger_rate}.  We recall that our predictions at [Fe/H]\,$\lesssim-2$ would be different
if we had used a more complex simulation with hydrodynamics or
cosmological structure formation (see discussions in Section~\ref{sect_pldtd}).
The numerical predictions with BH-NS mergers only (see middle panel of Figure~\ref{fig_gce_pop})
tend to drop near solar metallicity because of the lower frequency of black
holes at high metallicity (see references in Section~\ref{sect_ne_bl_pop_syn}).  On the other hand, except for the optimistic DTD functions, our GCE predictions
tend to increase near solar metallicity when using NS-NS mergers only (see upper panel of Figure~\ref{fig_gce_pop}), 
which is caused by a higher frequency of neutron stars.  This increasing trend
is not consistent with the observed monotonic decreasing trend of [Eu/Fe]
at [Fe/H]~$\gtrsim-1$.

Our GCE predictions show that it is difficult to explain the current content of Eu
in the Milky Way, at [Fe/H]\,$\sim0$, using the relatively low merger frequencies predicted by population synthesis
models.  This discrepancy is caused by
the normalization of the predicted DTD functions rather than by the shape of those functions, as
shown in Figure~\ref{fig_N_NSM_pop_GCE}.  We note that our
predicted [Eu/Fe] ratios could be enhanced by a 
factor of $\sim5$ if we had used the less iron-rich yield setup of other GCE
studies (see lower panel of Figure~\ref{fig_EuFe}).  Therefore we conclude 
that the GCE requirement and population synthesis predictions
can be in agreement, but only with the most optimistic merger rates, the 
largest Eu yields, and the lowest Fe yields. 

\section{Comparison with Advanced LIGO}
\label{sect_cmr}
As mentioned in the previous sections, the total number of NS-NS mergers needed
in GCE studies over a broad range of approaches are all consistent within a factor of $\sim4$.
This quantity, convolved with the cosmic star formation history, can be converted into a cosmic NS-NS merger rate
that can directly be compared with the current and upcoming Advanced LIGO measurements.

\subsection{Cosmic Star Formation History}
The cosmic star formation rate (CSFR) formula, as a function of redshift ($z$), has been taken from the recent study of
\citet{Madau2014},
\begin{multline}
\mbox{SFR}(z)=0.015\, {(1+z)^{2.7} \over 1+\left[(1+z)/2.9\right]^{5.6}} \\ \mathrm{[M}_\odot \, {\rm
Mpc}^{-3}\, {\rm yr}^{-1}\mathrm{]}.
\label{sfr}
\end{multline}
Due to the reddening and scarcity of good observational constraints, the CSFR is
not well established at $z>2$. The adopted formula is therefore very 
likely to represent a lower limit of the actual CSFR at such high redshifts. The adopted model does not fully 
correct for the small galaxies that were not measured in UV surveys, cannot account for the reionization of
the Universe at high-redshift, and underpredicts the observed gamma ray burst 
rate \citep{Kistler2009,Horiuchi2011,Mitchell2015}.   
Increasing the CSFR would increase the rate of double compact object mergers.  However, it
is not clear to which extent the CSFR should be increased. 
The adopted CSFR can be seen in Figure~1 of \cite{Belczynski2016c}.

\begin{figure}
\begin{center}
\includegraphics[width=3.25in]{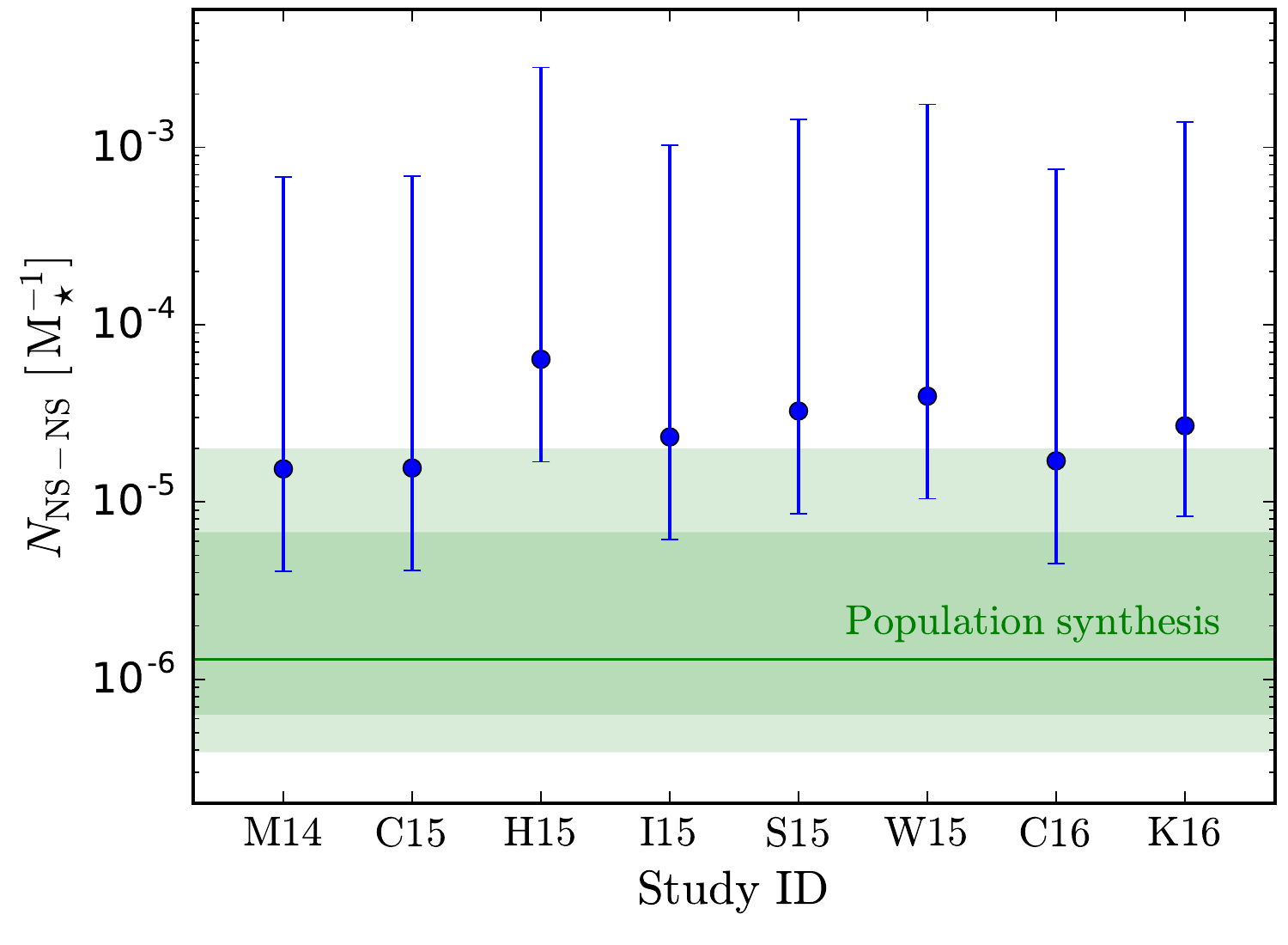}
\caption{Comparison between the normalized number of NS-NS mergers needed in CGE (blue) and the predictions from population synthesis models (green).  The solid green line represents the median value of the standard model, using equally spaced metallicities.  The dark green shaded area shows the range of values predicted by the standard model at different metallicities, while the light shaded area shows the same range once the two alternative models are included.  The blue circles represent the normalized values obtained using the fiducial ejected mass of Eu ($10^{-5}$\,M$_\odot$) per NS-NS merger event and the median ejected mass of Fe ($3.35\times10^{-4}$\,M$_\odot$) per stellar mass formed (see Figure~\ref{fig_EuFe}).  The blue vertical lines highlight the range of values that can predict the same level of [Eu/Fe] than the blue circles, which is defined by the plausible range of total masses ejected by NS-NS mergers (see Section~\ref{sect_ej_NSMs}) and by the variety of massive star Fe yields adopted in different GCE studies (see Figure~\ref{fig_EuFe}).}
\label{fig_N_NSM_pop_GCE}
\end{center}
\end{figure}

\subsection{Convolution Process}
To translate the NS-NS merger implementations found in GCE studies
into cosmic NS-NS merger rate densities, the CSFR is used as if it was the input
star formation history of a simple GCE model.  At each step (redshift bin),
a \textit{stellar population} is formed with a specific mass in units
of M$_\odot$\,Gpc$^{-3}$.  For each one of them, the evolution of the 
NMS rate is defined by the DTD function and its normalization. At any
redshift, the overall merger rate density is then obtained by summing the
contribution of all \textit{stellar populations}, which all have different ages.

\begin{figure*}
\begin{center}
\includegraphics[width=7in]{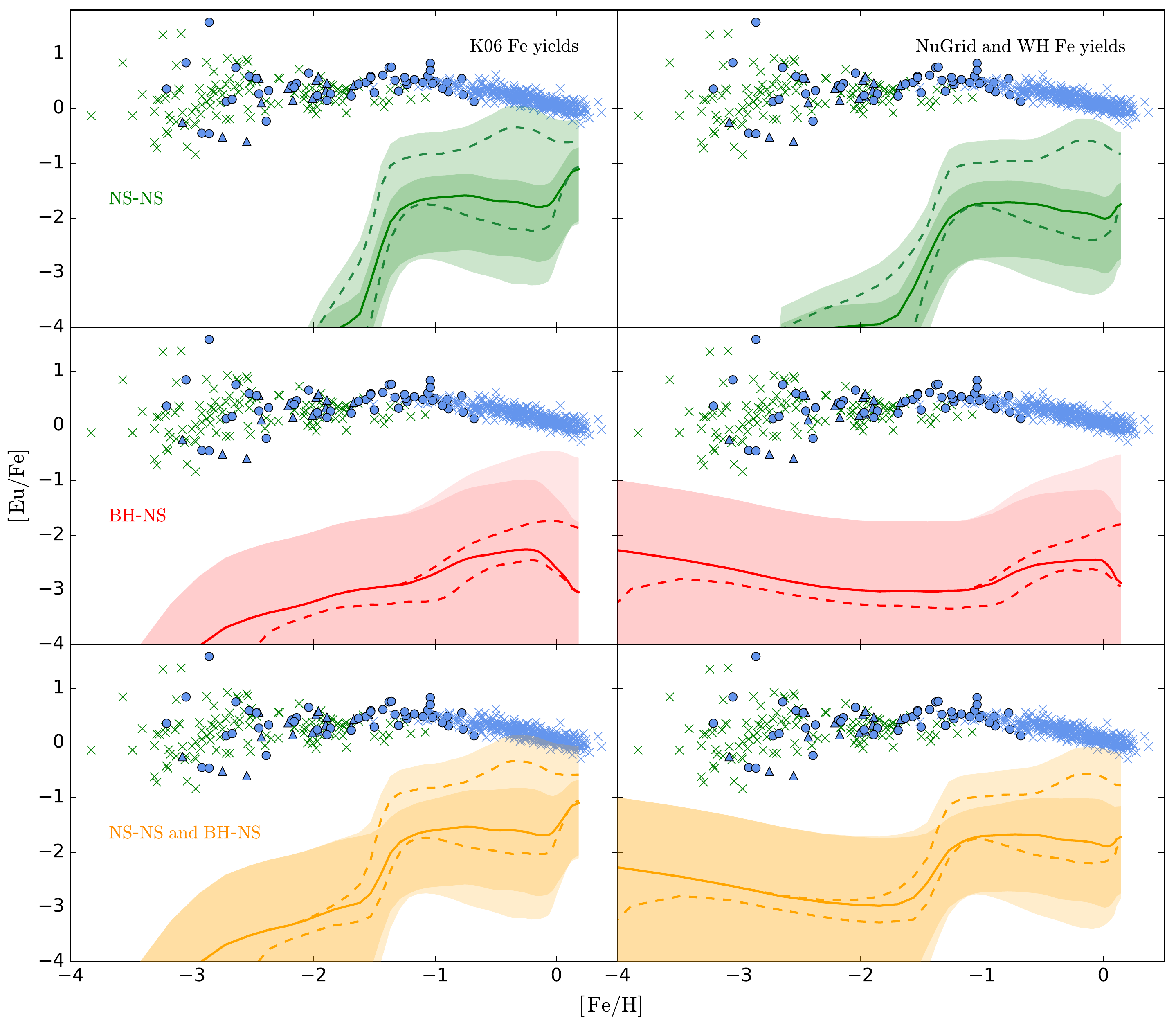}
\caption{Predicted chemical evolution of Eu using the metallicity-dependent population synthesis predictions of compact binary merger rates, combined with the Fe yields calculated by \citet[left panels]{k06} and by West \& Heger (in preparation) and NuGrid (right panels).  The upper, middle, and lower panels represent, respectively, the cases with NS-NS mergers only, with BH-NS mergers only, and with both NS-NS and BH-NS mergers.  We note that our IMF and yields setup typically generates more Fe than in other studies (see Figure~\protect\ref{fig_EuFe}).  Using alternative setups could then increase our predicted [Eu/Fe] levels by a factor of $\sim5$.  The solid lines represent the standard population synthesis models while the dashed lines represent the alternative models.  The dark and light shaded areas show the possible predictions defined by the wide range of masses ejected by NS-NS and by BH-NS mergers, for the standard model and the alternative models, respectively.  The observational data are the same as in Figure~\ref{fig_t_coal}.}
\label{fig_gce_pop}
\end{center}
\end{figure*}

\begin{figure*}
\begin{center}
\includegraphics[width=7in]{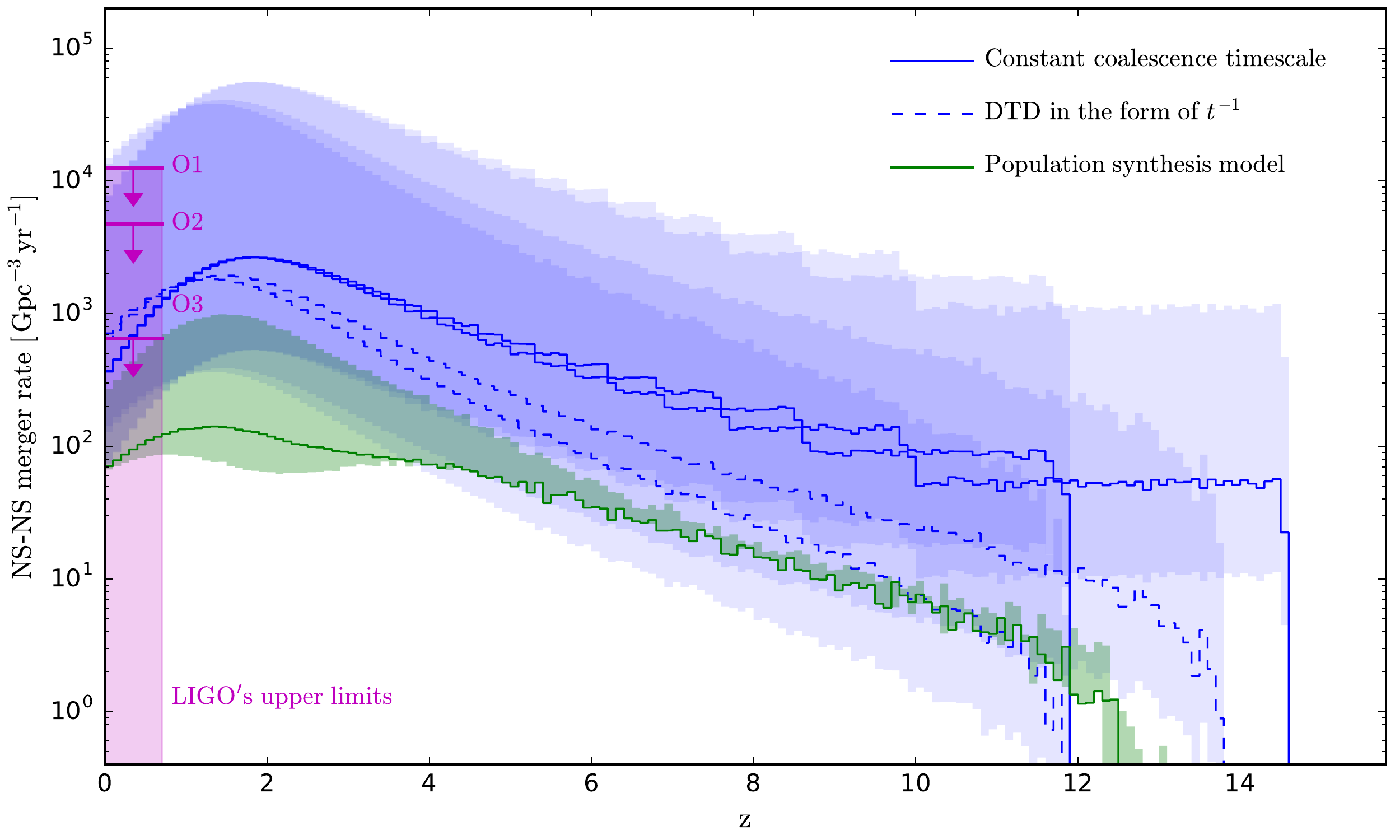}
\caption{Predicted cosmic NS-NS merger rate densities as a function of redshift.  The blue lines show the rate needed in galactic chemical evolution simulations, assuming different delay-time distribution functions (different line styles) and a total of $2\times10^{-5}$ NS-NS merger event per stellar mass formed.  The blue shaded areas highlight the possible range of values generated by using different ejected masses for NS-NS mergers and different Fe yields for massive stars (see Figure~\ref{fig_N_NSM_pop_GCE}).  The green line shows the predictions associated with the standard population synthesis model, while the green shaded area shows the possible range of values defined by the two alternative models.  The pink shaded area represents the upper limits established by Advanced LIGO during their first run of observations (\protect\citealt{aaa16b}).  O1 shows the current established value, while O2 and O3 are the expected values for the next observing runs.  The Advanced LIGO horizon goes up to $z\sim0.7$ for the most massive BH-BH mergers, but should be reduced for NS-NS and BH-NS mergers.}
\label{fig_cosmic_rate}
\end{center}
\end{figure*}

On the other hand, because our population synthesis models are
metallicity dependent, the convolution process into a cosmic framework 
necessitates an extra step.  The properties of compact binary objects first
need to be obtained as a function of redshift (see \citealt{Belczynski2016a}).
To do so, we assume that the evolution of the mean metallicity as a function
of redshift is given by the following formula, which is a modified version of the
one found in~\citet{Madau2014},
\begin{multline}
\log\big(Z_{\rm mean}(z)\big)=0.5\,+ \\ \log \left(  {y \, (1-R) \over \rho_{\rm b}}
\int_z^{20} {97.8\times10^{10} \, \mathrm{SFR}(z') \over H_0 \, E(z') \, (1+z')} dz' \right).
\label{Zmean}
\end{multline}
In this last equation, the fraction of the stellar mass ejected and returned into the interstellar
medium is $R=0.27$, the net metal yields of stellar ejecta is $y=0.019$, the 
baryon density is given by
\begin{equation}
\rho_{\rm b}=2.77 \times 10^{11} \,\Omega_{\rm b}\,h_0^2\,\mathrm{M}_\odot\,{\rm Mpc}^{-3},
\end{equation}
the CSFR is taken from equation~(\ref{sfr}), and
\begin{equation}
E(z)=\sqrt{\Omega_{\rm M}(1+z)^3+\Omega_{\rm k}(1+z)^2+\Omega_\Lambda}.
\end{equation}
The selected cosmological parameters are $\Omega_{\rm b}=0.045$, $\Omega_\Lambda=0.7$, $\Omega_{\rm M}=0.3$, $\Omega_{\rm k}=0$,
$h_0=0.7$, and $H_0=70\,{\rm km}\,{\rm s}^{-1}\,{\rm Mpc}^{-1}$.

We have increased the mean level of metallicity by $0.5$ dex at each
redshift to be in a better agreement with observational data~\citep{Vangioni2015}. 
At each redshift, we assume a log-normal distribution of metallicity around
the mean value with a dispersion of 0.5\,dex (see \citealt{Dvorkin2015}). 
This metallicity evolution is presented in the Figure~S2 of the supplementary
material of \cite{Belczynski2016b}.  We note that \cite{vangioni16} used the cosmic star formation
history along with population synthesis models to provide predictions for the evolution of r-process elements.

\subsection{Predicted Cosmic Merger Rates}
\label{sect_predcmr}
Figure~\ref{fig_cosmic_rate} shows the cosmic NS-NS merger
merger rates predicted by population synthesis models (green) and 
needed by GCE studies to reproduce the stellar abundances of Eu (blue).
The green shaded area represents the range of rates
defined by the alternative population synthesis models.
For the GCE studies, we assumed a total number of NS-NS mergers per
stellar mass formed of $2\times10^{-5}$, a coalescence timescales of
10 and 100\,Myr (solid blue lines), as in \blue{M14}, \blue{C15}, 
\blue{H15}, \blue{I15}, \blue{W15}, and Section~\ref{sect_cct}, and a DTD function in the
form of $t^{-1}$ (dashed blue lines) from 30 and 100\,Myr
to 10\,Gyr, as in \blue{S15}, \citealt{vv15}, \blue{K16}, and Section~\ref{sect_pldtd}.  The blue
shaded areas represent the range of rates that can
produce the same level of [Eu/Fe], which is defined by considering the
possible range of values for the total mass ejected by NS-NS mergers
and the mass of Fe ejected by massive stars (see blue lines in Figure~\ref{fig_N_NSM_pop_GCE}).  We note that the NS-NS frequency
adopted to calculate the predictions inferred by GCE studies (blue lines in Figure~\ref{fig_cosmic_rate})
is actually lower than what is needed in most GCE studies (see blue circles in Figure~\ref{fig_N_NSM_pop_GCE}).
The pink lines show the current (O1) and expected (O2, O3) upper limits
established by Advanced LIGO during their first run of observations (\citealt{aaa16b}).
Predictions derived from population synthesis models are well below the O3 upper limit,
while the ones derived from GCE studies overlap with all upper limits.  
Choosing between a constant coalescence timescale or a power law
for the DTD function of NS-NS mergers in GCE studies only changes the final cosmic
merger rates by a factor of two at most.

Using Figure~\ref{fig_cosmic_rate} as a reference, upcoming LIGO measurements 
will provide a constraint for both population synthesis models and GCE studies.
They will help to define whether or not NS-NS mergers can be the main source of r-process
elements in the Milky Way and its satellite galaxies.  If the measured cosmic NS-NS merger
rate, or its upper limit, drops below what is required in GCE studies when assuming NS-NS mergers
are the exclusive production sites of r-process elements, additional sites such as CC~SNe
are necessary (but see Section~\ref{sect_disc}).
In order to benefit from upcoming Advanced LIGO measurements
and to better quantify the contribution of NS-NS mergers,
the range of predictions associated with GCE studies (blue shaded areas in Figure~\ref{fig_cosmic_rate})
needs to be reduced.  We note that the thickness of the green shaded area associated with
population synthesis models should be considered as a lower limit, as a complete study
of plausible binary evolutionary scenarios still needs to be conducted (see Section~\ref{sect_sel_models}).

\begin{figure}
\begin{center}
\includegraphics[width=3.25in]{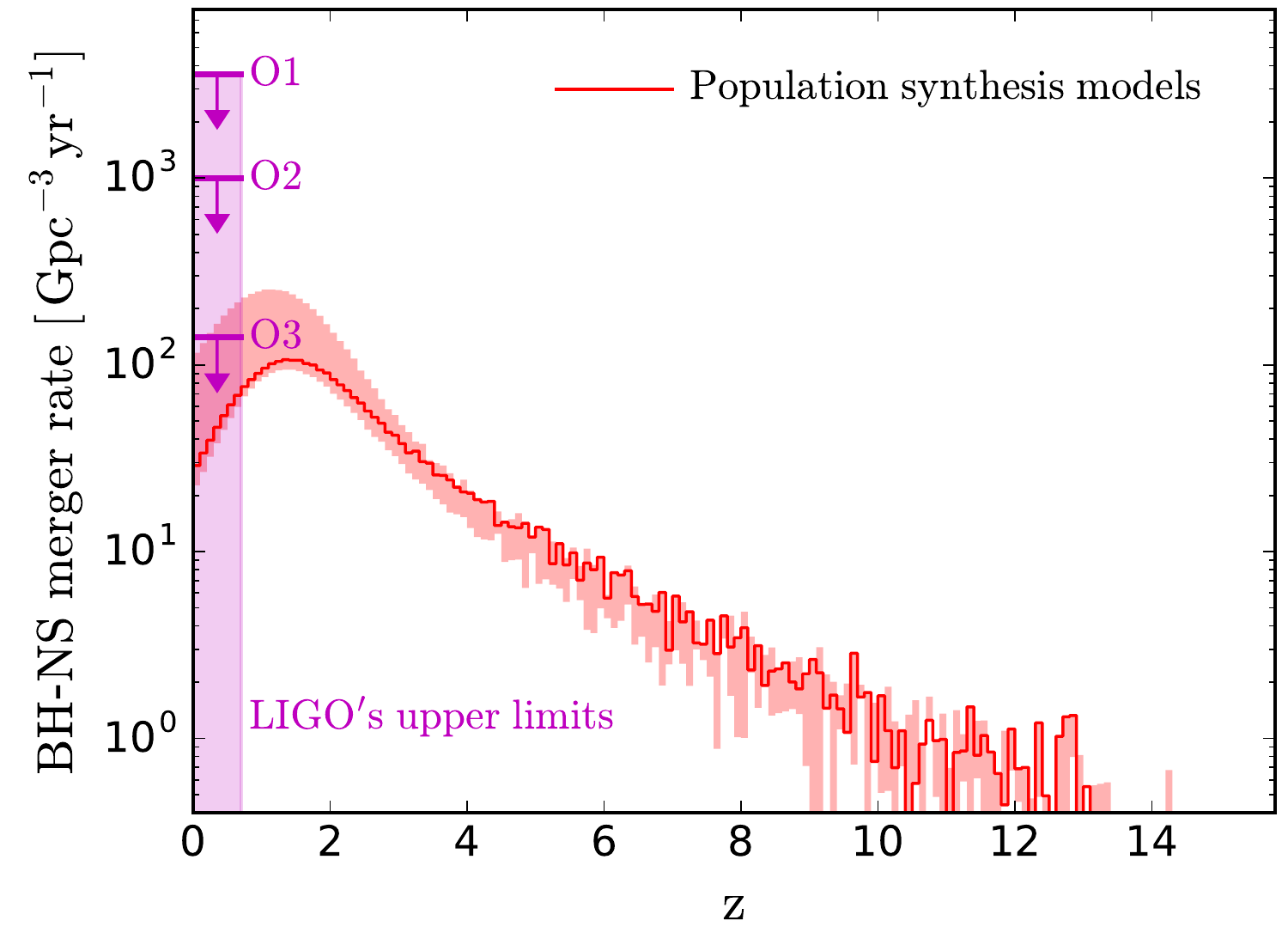}
\caption{Cosmic BH-NS merger rate as a function of redshift predicted by population synthesis models.  The lines and shaded areas are similar to the ones found in Figure~\ref{fig_cosmic_rate}.}
\label{fig_cosmic_bh_rate}
\end{center}
\end{figure}

The main assumption behind Figure~\ref{fig_cosmic_rate} is that the GCE requirement extracted from
Milky Way models and simulations is universal and representative of all galaxies that hosted
the NS-NS mergers that \textit{will} be visible inside the LIGO horizon (up to $z\sim0.1$).  For completeness, we show in Figure~\ref{fig_cosmic_bh_rate} the analogous of
Figure~\ref{fig_cosmic_rate} but for BH-NS mergers for population synthesis models.
As for NS-NS mergers, the predicted cosmic BH-NS merger rates are well below the O3 measured upper limit.

\section{Discussion}
\label{sect_disc}
In this section, we highlight additional sources of uncertainty that can affect our conclusions
and discuss the contribution of CC~SNe on the production of r-process elements.

\subsection{Type Ia Supernovae}
SNe~Ia are believed to be significant sources of Fe in the Milky Way (e.g., \citealt{mg86,f04,kob15}).
However, different GCE studies can adopt different SN~Ia implementations in terms of Fe yields (e.g., \citealt{t86,i99,travaglio04,s13}), total number of
explosions (e.g., \citealt{mmn14,c16b}), and delay-time distribution functions (e.g., \citealt{mr01,msfv09,y13}).  Those differences,
especially for the yields and total number of explosion, represent another source of uncertainty that can affect the
derived number of NS-NS mergers needed to reproduce the observed [Eu/Fe] vs [Fe/H] relationship.  
Considering a correction for SNe~Ia could improve the normalization process presented in Section~\ref{sect_nc} and
shown in Figure~\ref{fig_N_NSM}.

\subsection{Galactic Outflows}
\label{sect_disc_gal_out}
A significant source of uncertainty in modeling the early evolution of a
galaxy is the fraction of metals that is retained inside galaxies.  From a 
cosmological structure formation point of view, the high-redshift Universe should
be mainly composed of low-mass building block galaxies.  This implies
that galactic outflows should have been more important in the past,
as dwarf galaxies are believed to have stronger and more metal-rich
outflows than more massive systems (e.g., \citealt{t04,ps11,hop12,s12,m15}).
Even in the Local Universe, the majority of the metals that had been produced by stars seem to be
found outside galaxies (see the COS-Halo survey, \citealt{p14}).

Depending on the outflow history and on the production site for r-process elements, Eu and Fe
may not be returned into the interstellar medium on the same timescales
and therefore might not be retained inside galaxies in the same proportion.
On the other hand, the mass ejected from galaxies can also stay bounded to
the virialized system and eventually be reintroduced inside galaxies and 
recycled into stars.  Those gas exchanges between galaxies
and their circumgalactic medium could impact the chemical evolution of Eu,
depending on the different recycling timescales and on the fraction
of matter retained inside dark matter halos.

\subsection{Escaping Binaries}
Binary systems receive systemic velocities during the formation
of compact remnants both because of the mass lost during the explosions
and the kick imparted onto neutron stars and black holes at birth (see \citealt{fryer98} for a review).  
Population synthesis models of NS-NS and BH-NS systems show that
they have velocities ranging from $50$ to $1000{\rm \,km\,s^{-1}}$ (\citealt{fryer99}).
Because binary systems can sometime be accelerated
beyond the escape velocity of their host galaxy, 
some of the NS-NS and BH-NS mergers should occur outside galaxies (\citealt{fryer99,bloom99,bel06}).
Observations of short-duration gamma-ray bursts validated this model (\citealt{fb13}),
assuming NS-NS and BH-NS mergers are the progenitors of those bursts.
On the other hand, some fraction of compact binary systems have lower proper motions below
$\sim$\,30\,km/s (\citealt{kramer06,benia16}).
\cite{fryer98,fryer99} calibrated their NS kick distribution
to match the observed features of these low-velocity binary systems while also matching the pulsar
velocity distributions and the formation rates of X-ray binaries. This led to a bimodal kick distribution
and more low-velocity binaries compared to the single-mode kick distributions assumed
in this work (see Section~\ref{sect_psm}).

It is difficult at this point to derive the exact fraction of escape binaries.  Systems that received large
spatial velocities and escaped their birth sites could be harder to detect, which may induce an observational
bias.  But if compact binary systems can escape their host galaxies, and
are the main source of the r-process, this could imply that a fraction
of r-process elements can be found outside galaxies while the Fe
ejected by the progenitor stars could still be found inside galaxies.
This should induce an uncertainty in the derived number of compact binary
mergers needed in order to reproduce the [Eu/Fe] ratios in GCE studies.
This phenomena should be more important in the early
Universe where galaxies were smaller in size.  As a matter of fact,
\cite{bel06} suggested that, for small elliptical galaxies, the majority of
compact binary mergers occur outside their host galaxies.  We note
that massive stars can also be ejected outside galaxies by dynamical
interactions such as disruptions of binary systems or close encounters
with the galactic center.  However, as discussed in Section~\ref{sect_disc_gal_out},
the Eu ejected outside galaxies could still be reintroduced inside galaxies
after a certain time, depending on the dark matter gravitational potential well of the host
galaxy when escaping binary mergers occur.

\subsection{Stellar Remnants}
The stellar remnant mass distribution plays a crucial role in both
chemical evolution and population synthesis models, as it determines
the relative fraction of black holes and neutron stars.
Several observations suggest that the high-mass end of the massive
star spectrum should preferentially form black holes -- the lack
of CC~SN progenitor stars with initial mass above 18\,M$_\odot$
(\citealt{smartt15}), the black hole mass distribution function (e.g., \citealt{bwf12}),
the failed explosion of a 25\,M$_\odot$ star (\citealt{a16}), and the
merger of two black holes of $\sim$\,30\,M$_\odot$(\citealt{aaa16}).
A similar conclusion is motivated by theory (e.g., \citealt{f99,heger03})
but can be more complex with predictions of islands of non-explodability
(\citealt{u12,ertl16,suk16}) and of highly non-monotonic relations between
the stellar initial and final masses resulting from binary interactions (e.g, \citealt{Belczynski2008c}).

For GCE applications, accounting for direct black hole formation generates different
chemical evolution predictions compared to when all massive stars are assumed
to produce a CC~SN (see \citealt{c16c}).  When stellar yields use a relatively
flat and low remnant mass distribution (e.g., \blue{K06}; \citealt{cl13}), the ejected masses of alpha and
iron-peak elements stay roughly constant or continuously increase as a function
of stellar initial mass.  When using stellar yields that account for black hole
formation (e.g., \citealt{fbw12,p13}; C. Ritter et al. in preparation), the ejected mass
for all metals usually start to decrease in stars more massive than $\sim$\,25\,M$_\odot$.

Depending on the set of yields and the adopted remnant mass distribution,
predictions for the evolution of alpha and iron-peak elements, including Fe,
can significantly differ and can thus modify the number of NS-NS and BH-NS mergers
needed in GCE studies to reproduce the observed [Eu/Fe] abundance ratios.
In addition varying the remnant mass distribution in stellar populations can also modify
the overall rates of NS-NS and BH-NS mergers.  Because these two types
of binary mergers do not necessarily produce the same ejecta
(see Section~\ref{sect_me}), the choice of the remnant mass distribution 
can therefore have an impact on the total mass of r-process elements
returned into the interstellar medium and recycled into stars.

\subsection{Core-Collapse Supernovae}
In this paper, compact binary mergers have been considered as the exclusive production sites of r-process elements, as our goal was to investigate whether or not NS-NS mergers alone can reproduce the current amount of Eu observed in the stars of the Milky Way.  The relative contribution, especially at low [Fe/H], of NS-NS mergers is still highly debated and CC~SNe could still be important production sites.  \blue{M14} suggested an early enrichment by CC~SNe originating from progenitor stars in the mass range of 12\,--\,30\,M$_{\odot}$, 20\,--\,50\,M$_{\odot}$, or a combination of these, which is also suggested by \blue{C15}.  Another production site of r-process elements are the magnetorotationally driven SNe (MHD~SNe, e.g., \citealt{Winteler12,nishimura15}).  These rare SNe generate highly magnetized and fast rotating neutron stars that represent $\lesssim1$\,\% of all neutron stars.  \blue{C15} and \blue{W15} adopted a combined environment which considers the contributions of both NS-NS mergers and MHD~SNe as r-process production sites and found that such combinations are able to explain the observed r-process abundances, especially at low [Fe/H].

Although we only considered compact binary mergers in our study, it should be noted that other production sites can still contribute significantly to the evolution of the r process.  Our appellation of \textit{how many NS-NS mergers are needed in GCE studies} should be considered as an upper limit of the actual contribution of such events.

\section{Summary and Conclusions}
\label{sect_concl}
A connection between GCE, population synthesis models, and Advanced LIGO measurements have been made in order to investigate whether or not NS-NS mergers (and compact binary mergers in general) can be the main source of r-process elements in the Milky Way.  We have compiled and reviewed eight GCE studies, including our own, and extracted how many NS-NS mergers are needed in order to reproduce the evolution of Eu in our Galaxy.  Those studies include a wide rage of numerical setup from simple one-zone homogeneous-mixing models to cosmological hydrodynamic simulations.  In spite of the different levels of complexity of the considered studies, we found that, once normalized, the needed number of NS-NS mergers converges within a factor of $\sim4$ (but see Section~\ref{sect_disc}).

Using {\tt OMEGA}, our one-zone GCE code, we explored the impact of using different DTD functions for NS-NS mergers on our ability to reproduce the Eu abundances trend.  Using a constant coalescence timescale, after which all NS-NS mergers occur in a simple stellar population, our predictions regarding the evolution of Eu are consistent with previous studies that used a similar implementation.  We cannot reproduce, however, observations at low [Fe/H] when using long-lasting DTD functions in the form of a power law with an index of $-1$ or lower.  This highlights the inability of simple models to capture the early evolution of our Galaxy, as more complex simulations that include hydrodynamics (\blue{S15}; \blue{H15}; \citealt{vv15}) and a proper treatment of cosmological structure formation (\blue{K16}) can reproduce, using NS-NS mergers only, the Eu abundances of metal-poor halo stars, even with long-lasting DTD functions.  Nevertheless, the overall number of NS-NS mergers needed to reach [Eu/Fe]~$\sim0$ at [Fe/H]~$\sim0$ is relatively insensitive of the choice of the DTD function.

We compared the number of NS-NS mergers needed in GCE studies with the predictions of population synthesis models, accounting for the wide range of total masses ejected by NS-NS mergers and Fe yields associated with massive stars.  We then introduced the metallicity-dependent NS-NS and BH-NS mergers rates predicted by population synthesis models into our GCE model as an input to see their impact in a GCE context.  Finally, we convolved the GCE needed NS-NS merger implementations and population synthesis predictions with the cosmic star formation history to provide results that can directly be compared with upcoming Advanced LIGO measurements.  Our main conclusions are the following:

\begin{itemize}
\item{GCE typically requires about 10 times more NS-NS mergers than what is predicted by our standard population synthesis models, when assuming that the ejected r-process mass is 0.01\,M$_\odot$ per merger event.}
\item{It is difficult to reproduce Eu observations by introducing the metallicity-dependent populations synthesis predicted rates into our GCE model, even when considering both NS-NS and BH-NS mergers.}
\item{GCE and population synthesis can only be in agreement when assuming optimistic rates, high Eu yields, and low Fe yields.}
\item{The cosmic NS-NS and BH-NS merger rate densities predicted using population synthesis models are below all upper limits currently established by Advanced LIGO.  The cosmic NS-NS merger rate densities
inferred using the GCE requirement are uncertain by a factor of $\sim150$ and overlap with all upper limits.}
\item{Our population synthesis models suggest that BH-NS mergers appear earlier (at lower [Fe/H]) than NS-NS mergers but overall,
contribute less to the evolution of the r process.  This is under the assumption that the maximum ejected
mass is 0.1\,M$_\odot$ per merger event and that no more than half of BH-NS mergers can eject material (see Section~\ref{sect_ej_BH_NSMs}).}
\end{itemize}

At the moment, population synthesis models and GCE studies are in agreement 
with the O1 upper limit currently established by Advanced LIGO. 
Using Figure~\ref{fig_cosmic_rate} as a future reference, Upcoming Advanced LIGO
measurements will provide valuable insights on the plausibility of the GCE requirement
and will help to define whether or not NS-NS mergers can be the main source of the r-process elements.
We recall that our conclusions are based on the assumption that NS-NS mergers should not,
on average, eject more than 0.025\,M$_\odot$ of material per merger event (see Section~\ref{sect_ej_NSMs}).
Furthermore, we recall that our alternative population synthesis models
may not be representative of the complete plausible range of NS-NS and BH-NS merger
rates.  More calculations are needed in order to isolate extreme cases that will
provide more realistic evolutionary uncertainties.

Besides the detection of gravitational waves that can constrain the rate of compact binary mergers,
kilonovae could constrain the mass ejected by NS-NS mergers.  However,
current uncertainties in the opacities (\citealt{fontes15}) make it difficult to extract the r-process ejecta from a
kilonova observation.  Both detailed observations and high-fidelity models are needed.  But if they can
be done, these detections could be use to constrain the r-process ejected mass.

Many sources of uncertainty, such as the yields ejected by compact binary mergers,
the binary escape fraction, the circulation of gas inside and outside
galaxies, the cosmic star formation history, and different SN~Ia implementations, can still affect the
reliability of our conclusions.  Ongoing work in nuclear
astrophysics, in observation, and in galaxy evolution will hopefully contribute to reduce the current
level of uncertainty, which will lead to a better quantification of the role played by compact
binary mergers in the evolution of r-process elements.

\section*{Acknowledgements} 
We are thankful to Yutaka Hirai and Yutaka Komiya for sharing details of
their chemical evolution codes, to Marco Pignatari for providing useful
analysis on NuGrid stellar models, to Christopher West and Alexander
Heger for providing unpublished stellar yields, and to Oleg Korobkin for
discussions on the r process.  This research is supported by the National
Science Foundation (USA) under Grant No. PHY-1430152 (JINA Center
for the Evolution of the Elements), and by the FRQNT (Quebec, Canada)
postdoctoral fellowship program.  KB acknowledges support from the NCN grants Sonata Bis 2
(DEC-2012/07/E/ST9/01360), OPUS (2015/19/B/ST9/01099) and 
OPUS (2015/19/B/ST9/03188).  CLF is funded in part under the auspices of
the U.S. Department of Energy, and supported by its contract W-7405-ENG-36
to Los Alamos National Laboratory.  BW is supported by the European
Research Council (FP7) under ERC Advanced Grant Agreement
No. 321263 - FISH, and the Swiss National Science Foundation (SNF).
The Basel group is a member in the COST Action New CompStar.
BWO was supported by the National Aeronautics and Space Administration
(USA) through grant NNX12AC98G and Hubble Theory Grant HST-AR-13261.01-A.

\appendix
\section{Highly Eccentric Binaries}
\label{ap_1}
\begin{figure}
\begin{center}
\includegraphics[width=4.in]{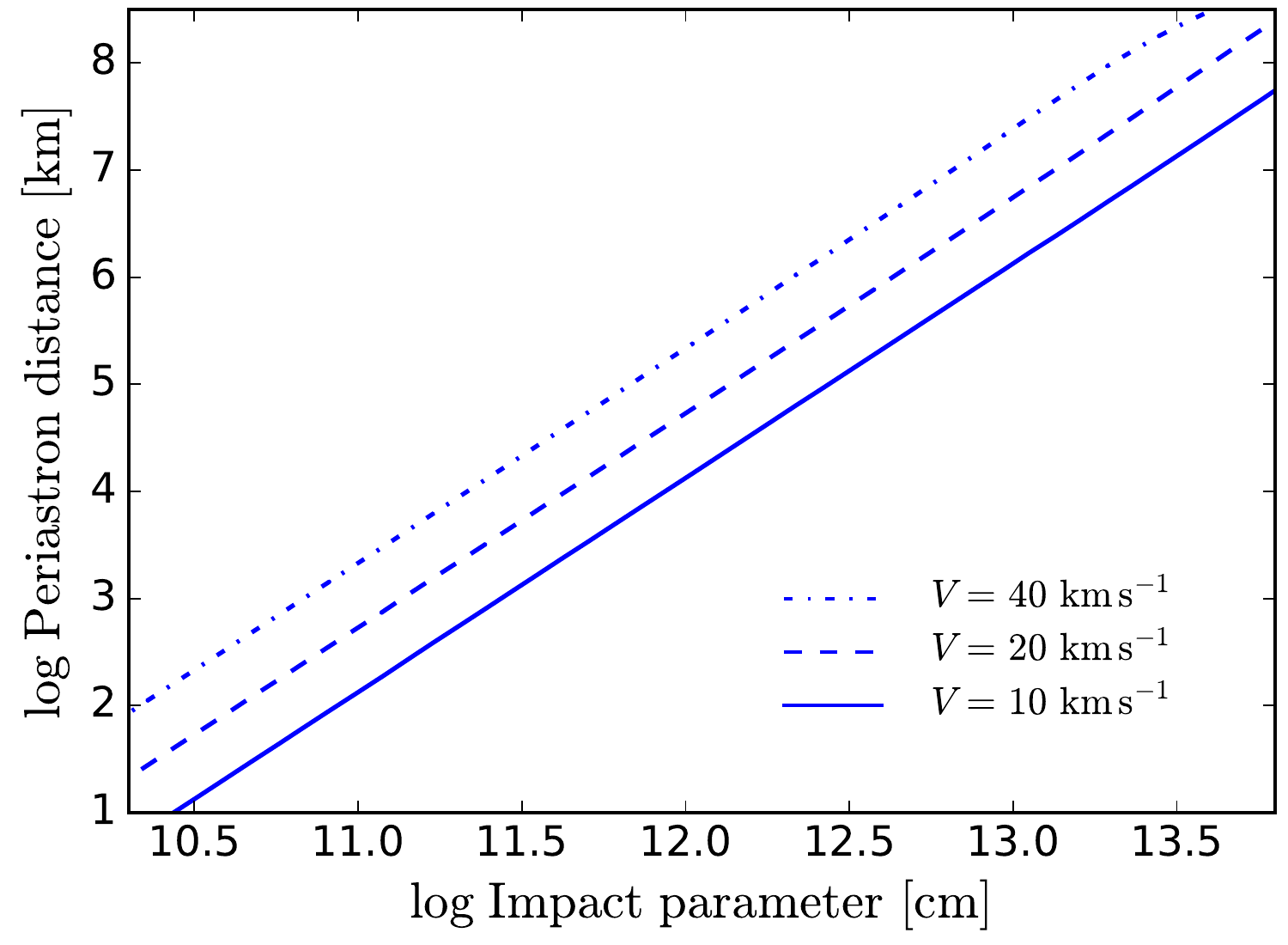}
\caption{Periastron distance as a function of the impact parameter for different impact velocities.}
\label{fig_avr}
\end{center}
\end{figure}

We used a dynamics code to
model the periastron distances of compact binary systems for a range of impact parameters and velocities.
As an example, for two 1.4\,M$_\odot$ neutron stars interacting with an impact velocity of 10\,km\,s$^{-1}$,
the system remains bound if the impact parameter is $10^{14}$\,cm.  But for such wide separations,
the resulting periastron is above $10^{8}$\,km (see Figure~\ref{fig_avr}).  Only when the impact parameter
is below $10^{11}$\,cm does the periastron distance drops below 20\,km, which would allow NS-NS mergers
to eject more than 0.025\,M$_\odot$ of r-process material (see \citealt{radice16}).

The cross-section of these
interactions is proportional to the impact parameter squared and we can determine the relative rates between
collisions and mergers from dynamical interactions by taking the ratio of these cross sections.  Assuming we
need periastron values below 20\,km, the frequency ratio of collisions relative to mergers is about $4\times10^{-8}$.
Figure~\ref{fig_avr} shows the periastron separation as a function of the impact parameter for different values
of impact velocities.  The rate of eccentric neutron star binaries (or collisions) is extremely rare
and, overall, these systems are not expected to contribute significantly to the production of r-process material.

\label{lastpage}

\end{document}